\documentclass[12pt,aps,prb,onecolumn,noshowpacs]{revtex4-1}
\usepackage{amsmath,graphicx,amsbsy,amssymb,amsmath,epsfig,latexsym,wasysym,pifont,mathrsfs,natbib}
\usepackage{float} \newfloat{widefig}{thp}{lop} \usepackage{comment}
\usepackage{array, makecell} \usepackage{verbatim} \usepackage{datetime}
\usepackage{multirow,array} \usepackage{hyperref} \usepackage{color} \usepackage{setspace}
\usepackage{subcaption}
\usepackage{graphicx}
\usepackage{dirtytalk}
\usepackage[british]{babel}
\usepackage{csquotes}
\usepackage{comment}
\usepackage{appendix}
\usepackage[utf8,latin1]{inputenc}
\usepackage[export]{adjustbox}
\UseRawInputEncoding

\hypersetup{colorlinks,
  linkcolor=blue,%
  citecolor=blue,%
  urlcolor=blue}

\begin{document}

\title{\textbf{Effects of Chemical and magnetic disorder on the electrochemical properties of V$_{2-x}$Mn$_{x}$CO$_{2}$} MXene electrodes}

\author{Mandira Das}
\affiliation{Department of Physics, Indian Institute of Technology
  Guwahati, Guwahati-781039, Assam, India.}  
\author{Himanshu Murari}
\affiliation{Department of Physics, Indian Institute of Technology
  Guwahati, Guwahati-781039, Assam, India.}  
  \author{Biplab Sanyal}
 \affiliation{Department of Physics and Astronomy,
  Uppsala University, Sweden} 
\author{Subhradip Ghosh}
\email{subhra@iitg.ac.in} \affiliation{Department of Physics,
  Indian Institute of Technology Guwahati, Guwahati-781039, Assam,
    India.} 
\begin{abstract}  
Investigation of structure-property relations in chemically and magnetically disordered materials can give rise to interesting physical phenomena. The potential of two-dimensional MXenes as electrodes in supercapacitor applications have been studied extensively. However, the role of chemical and magnetic disorder on their electrochemical parameters like the capacitance have not been explored yet. In this work, we have systematically addressed this for V$_{2-x}$Mn$_{x}$CO$_{2}$ MXene solid solutions with an analysis based upon results from first-principles electronic structure calculations. We find that the variations in the total capacitance over a voltage window depends upon the degree of chemical and magnetic disorder. In course of our investigation, we also found out that the magnetic structure on the surface can substantially influence the redox charge transfer, an yet unexplored phenomenon. A significantly large charge transfer and thus a large capacitance can be obtained by manipulating the chemical composition and the magnetic order of the surfaces.These findings can be useful in designing operational supercapacitor electrodes with magnetic constituents.  
\end{abstract}


\maketitle

\section{Introduction\label{intro}}
The excitement of materials researchers over the discovery of two-dimensional (2D) materials with novel properties amplified after the discovery of MXenes, a class of layered compounds formed by early transition metals and carbon or nitrogen. Since the discovery of Ti$_3$C$_2$, the first MXene, in 2011\cite{naguib2011two}, it has been explored extensively for potential applications in multiple areas like energy material\cite{lukatskaya2013cation,mashtalir2013intercalation}, electromagnetic interference shielding\cite{han2020beyond}, memristors\cite{yan2019new}, gas sensors\cite{pei2021ti3c2tx}, optical sensors\cite{pei2021ti3c2tx}, photocatalysts\cite{li2021applications}. The promising performances of Ti$_3$C$_2$ in various applications propelled researchers to look for other MXenes. Accordingly, several other MXenes have either been synthesized \cite{lukatskaya2013cation,hu2016high,chandran2020mos2,du2018mxene,liu2020ti3c2tx}  or predicted to form as per sophisticated computational modeling \cite{hong2020double}. Though Ti$_3$C$_2$T$_x$ MXene turned out to be the most investigated one in the family, in recent times variety of MXenes with single transition metal component (V$_2$CT$_x$\cite{he2020two,tan2021v}, Ti$_2$CT$_x$\cite{zhu2019synthesis,shi2022vertical}, Nb$_2$CT$_x$\cite{xiao2021one,byeon2016lithium}, for example), with double transition metal components (Mo$_2$TiC$_2$T$_x$)\cite{anasori2015two,cheng2018two}, and with a fixed concentration of the transition metal constituents (i-MXenes Mo$_{1.33}$C\cite{el2021enhanced}, Nb$_{1.33}$C\cite{ahmed2020mxenes,doi:10.1021/acsanm.8b00332}) have been investigated. 
Solid solutions of compounds are often considered to be the route to achieve improvements in the target properties. Attempts in this direction have been made for MXenes as well. Consequently, (Ti,V)$_3$C$_2$\cite{shen2018novel}, (Nb,V)$_2$C\cite{wang2021adjustable,han2020tailoring} and (Ti, Nb)$_2$C,\cite{wang2021adjustable,han2020tailoring} (Ti,V)$_2$C\cite{han2020tailoring} MXene solid solutions have already been synthesized. The variations in the electrochemical performances of Ti$_{2-y}$Nb$_{y}$T$_x$ and V$_{2-y}$Nb$_y$T$_x$ ($0<y<2$) solid solutions as a function of compositions have been investigated by Wang $et.al.$\cite{wang2021adjustable}. To our knowledge, this is the first comprehensive study on the electrochemical properties of solid solution MXene. In this work, they found that the charge storage capacities of (Ti,Nb)$_2$C and (V,Nb)$_2$C solid solutions lie between the end-point compounds. Apart from these, several solid solutions of MAX compounds, the precursor to MXenes, have been prepared \cite{hong2020double}. The synthesis of MAX compounds implies that obtaining MXene solid solutions with those constituents is only a matter of time. 
Prominent among the solid solutions in MAX compounds are (V, Mn)$_2$C, (Cr, V)$_2$C, and (Cr, Mn)$_2$C \cite{hong2020double}. These compounds stand out as either one or both transition metal constituents are magnetic elements. Although the study of magnetism in 2D materials is still in its infancy, enormous potential in the fabrication of devices by exploiting different magnetic orders makes them interesting research subjects. The long-range orders found in recently synthesised 2D CrI$_3$ \cite{huang2017layer}, VSe$_2$ \cite{wu2023effects}, Cr$_2$Ge$_2$Te$_6$ \cite{song2019surface} and Fe$_3$GeTe$_2$ \cite{fei2018two} have given the research on exploring magnetism in 2D crystals a filip. In recent years, investigations into the structure-magnetic property relations in MXenes have also started. Although the number of experimental results on magnetic MXenes, to date, is limited, a plethora of first-principles electronic structure calculations \cite{si2015half,sun2021cr2nx2,yue2021tuning,he2016new,zhang2022computational} have revealed exciting magnetic behavior in them. 
With regard to studies on MXenes in assessing their potential for various applications, what is absent is studies on the impact of magnetism on properties other than the magnetic ones, for example, the electrochemical performances. As MXenes have mostly been investigated with regard to their electrochemical performances so that they can be used as suitable electrodes in electrochemical reactions, it is only appropriate to assess the electrochemical performances of MXenes having one or more magnetic constituents. The role of magnetic orders on the electrochemical performances in such systems would provide important insights into the interrelations of magnetic and chemical interactions. Moreover, the simultaneous presence of chemical and magnetic disorders in a single material can give rise to interesting and novel phenomena that can provide a different perspective on the physics and chemistry of the material. Solid solution MXenes with magnetic components are tailor-made systems for such investigations into the effects of chemical and magnetic disorders on their electrochemical properties.
Using Density Functional Theory (DFT)\cite{dft} based electronic structure methods, in this work, we investigate the impacts of chemical and magnetic orders on the electrical and geometrical contributions to the total capacitances and, thus, the charge storage capacities of oxygen (-O) functionalized V$_{2-x}$Mn$_x$C MXenes. We systematically investigate the effects of different disorders by tuning the degree of spin disorder for systems with a given chemical composition. Our work, for the first time, addresses such issues by making an in-depth exploration of physics at the microscopic level. Our results illustrate the inter-dependence of chemical and magnetic disorders in modifying the charge storage capacity of MXenes, an important aspect from the point of view of device fabrication. 
\section{Methodology and Calculational Details\label{methods}}
\subsection{Modeling the solid-solution structures \label{models-structure}}
\begin{figure}[H]
    \centering
    \begin{subfigure}[b]{0.45\linewidth}
    \includegraphics[width=\linewidth]{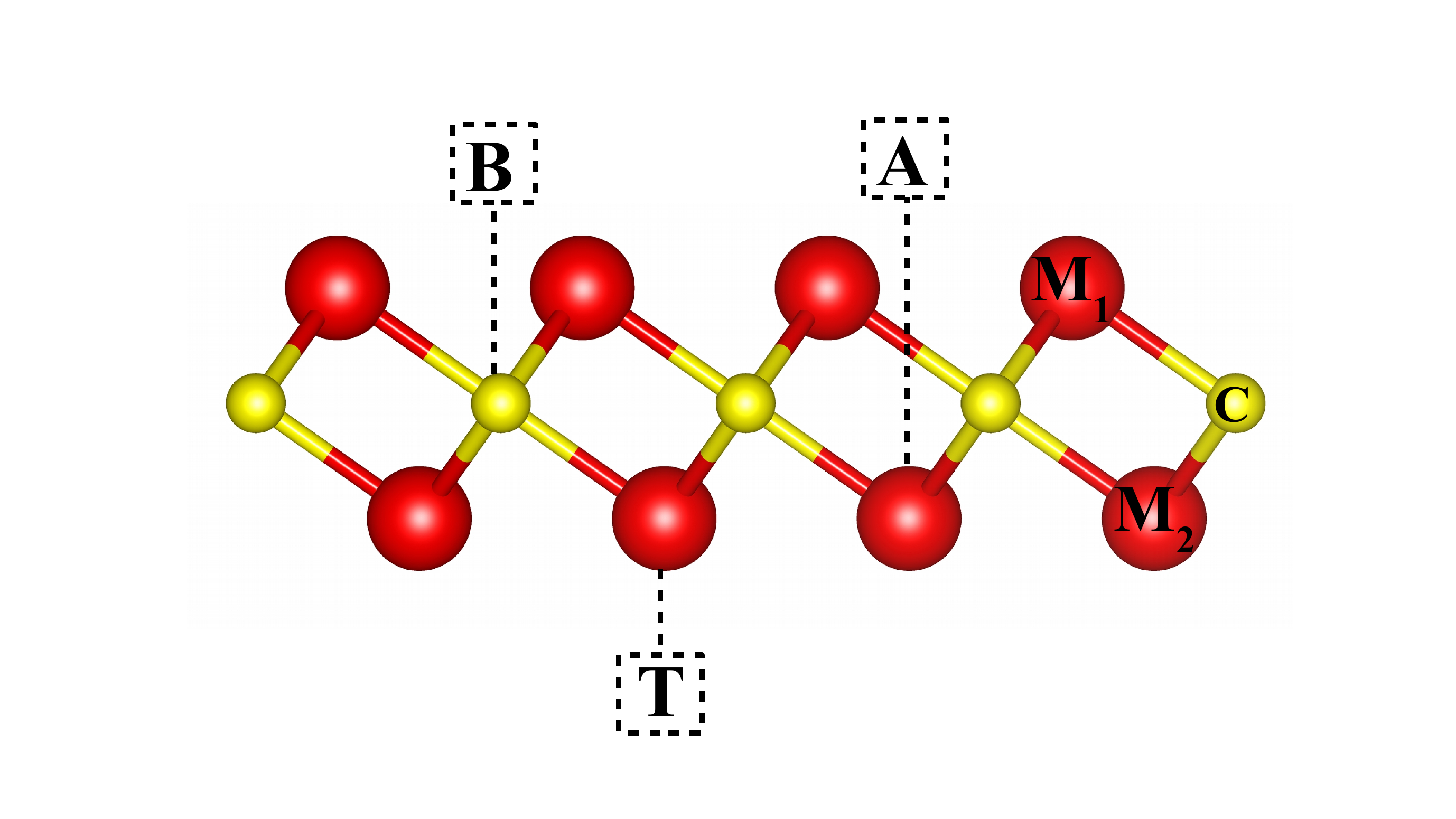}
    \caption{Side view}
    \end{subfigure}
    \hspace{-0.05cm}
    \begin{subfigure}[b]{0.45\linewidth}
    \includegraphics[width=\linewidth]{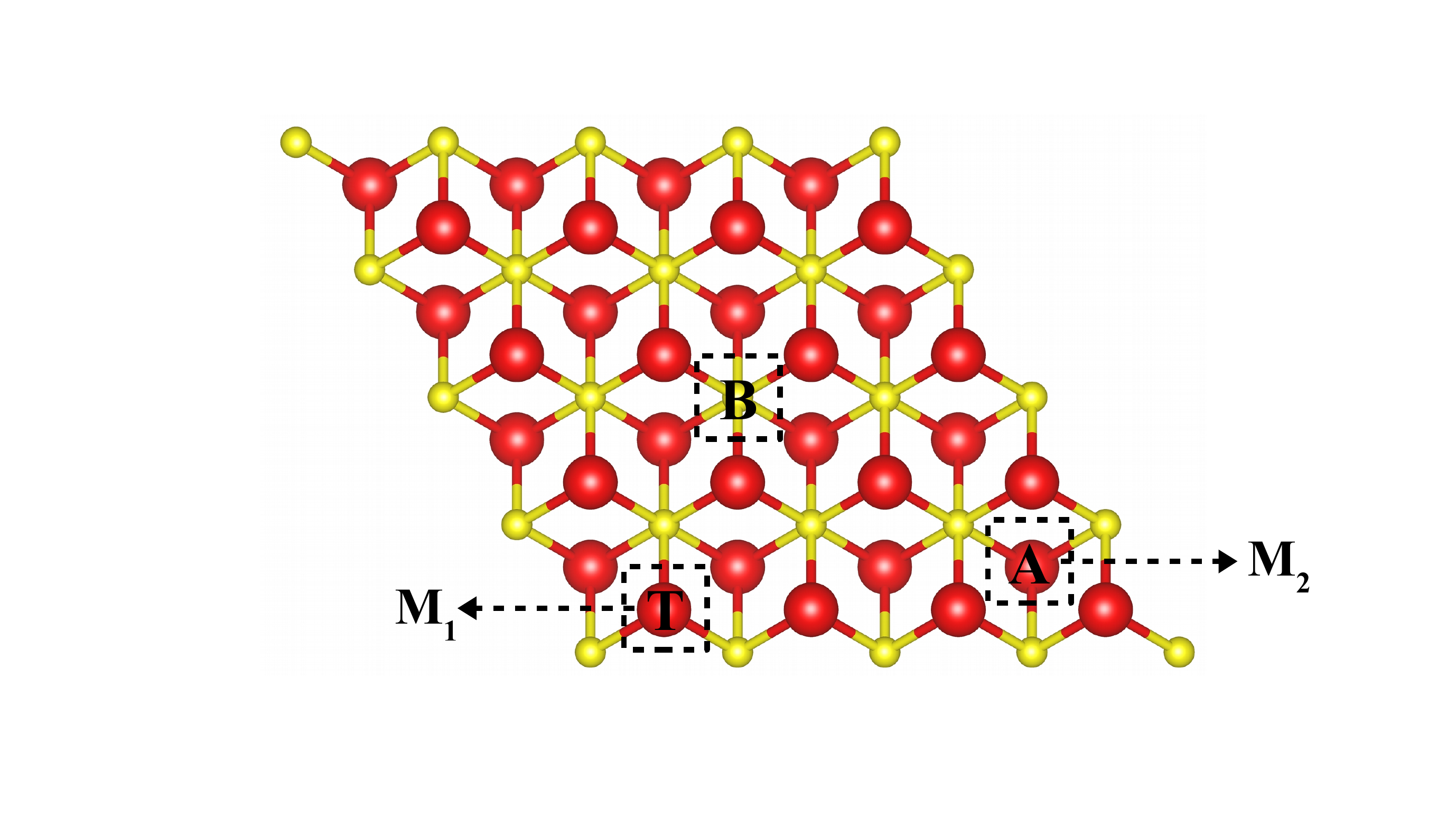} 
    \caption{Top view}
    \end{subfigure}
    \caption{Illustrative representation of different oxygen passivation sites in M$_2$C MXene: (a) side view and (b) top view . Red and yellow colored balls represent M and C atoms, respectively. M$_1$ and M$_2$ are the top and bottom layer M elements. }
    \label{Fig:1}
\end{figure}

MXenes with chemical formula M$_{n+1}$X$_{n}$ (M the transition metal, X Carbon or Nitrogen and $n$ a non-zero positive integer) are obtained from corresponding MAX compounds M$_{n+1}$AX$_{n}$ by removing the \textquote{A} layers. The resulting structure is of trilayer sheets arranged in a hexagonal-like unit cell (P6$_{3}$/mmc, space group 194) where the X layers are sandwiched between two M layers. Our modeling of the 2D solid-solution structure of V$_{2-x}$Mn$_x$C; $x=0.5,1,1.5$ starts by the construction of a 4$\times$4$\times$1 cell of optimized V$_2$C MXene. The monolayer is modeled by keeping a large inter-layer distance of 15$\AA$. The solid solutions that are completely random alloying of V and Mn are then obtained by replacing V with Mn atoms in such a way that each Mn has the same environment. This results in 12, 12, and 4 configurations for systems with $x=0.5,1.5$ and $1$, respectively. The stabilities of these configurations are assessed by computing their formation energies
\begin{equation}
     E_{formation} = E(V_{2-x}Mn_{x}C)- [(2-x)E(V) + (x)E(Mn) + E(C)]
    \label{EQN: 1}
\end{equation}
$E(V_{2-x}Mn_{x}C);x=0.5,1.0,1.5$, is the total energy of the solid solution considered, and $E(Mn), E(V)$, and $E(C)$ are the energies of constituent elements in their respective ground states. 
The process of exfoliation of MXene from the MAX phase inevitably results in surface passivation of by -F, -O, and -OH functional groups\cite{naguib2011two}. There is enough evidence to suggest that the -O functional group primarily contributes to the redox capacitance, whereas -F and -OH contribute to the electrical double-layer capacitance\cite{zhan2018understanding}. For the purpose of our investigation, in this paper, We consider -O surface passivated solid solution MXenes V$_{2-x}$Mn$_{x}$CO$_{2}$. To ascertain the positions of -O on the MXene surface, we consider models of O-functionalized MXene\cite{struct}. The possible positions of O atoms on the surfaces are shown in Figure \ref{Fig:1}.
There are two hollow sites available for the O atoms to occupy: (a) above or below the transition metal(M) and on the distant surface, marked as A, (b) above and below the C atom, marked as B in Figure \ref{Fig:1}. The position right on top of the $M$ site (marked as T in Figure \ref{Fig:1})is also a possible site for the O atom to occupy. Accordingly, six different O occupancy models are possible. In this work, we have considered all six models for finding the most stable configuration of the functionalized MXene solid solutions.

\subsection{Modeling of the Magnetic structures \label{models-magnetic}}
\begin{figure}[h]
    \includegraphics[width=0.50\textwidth,center]{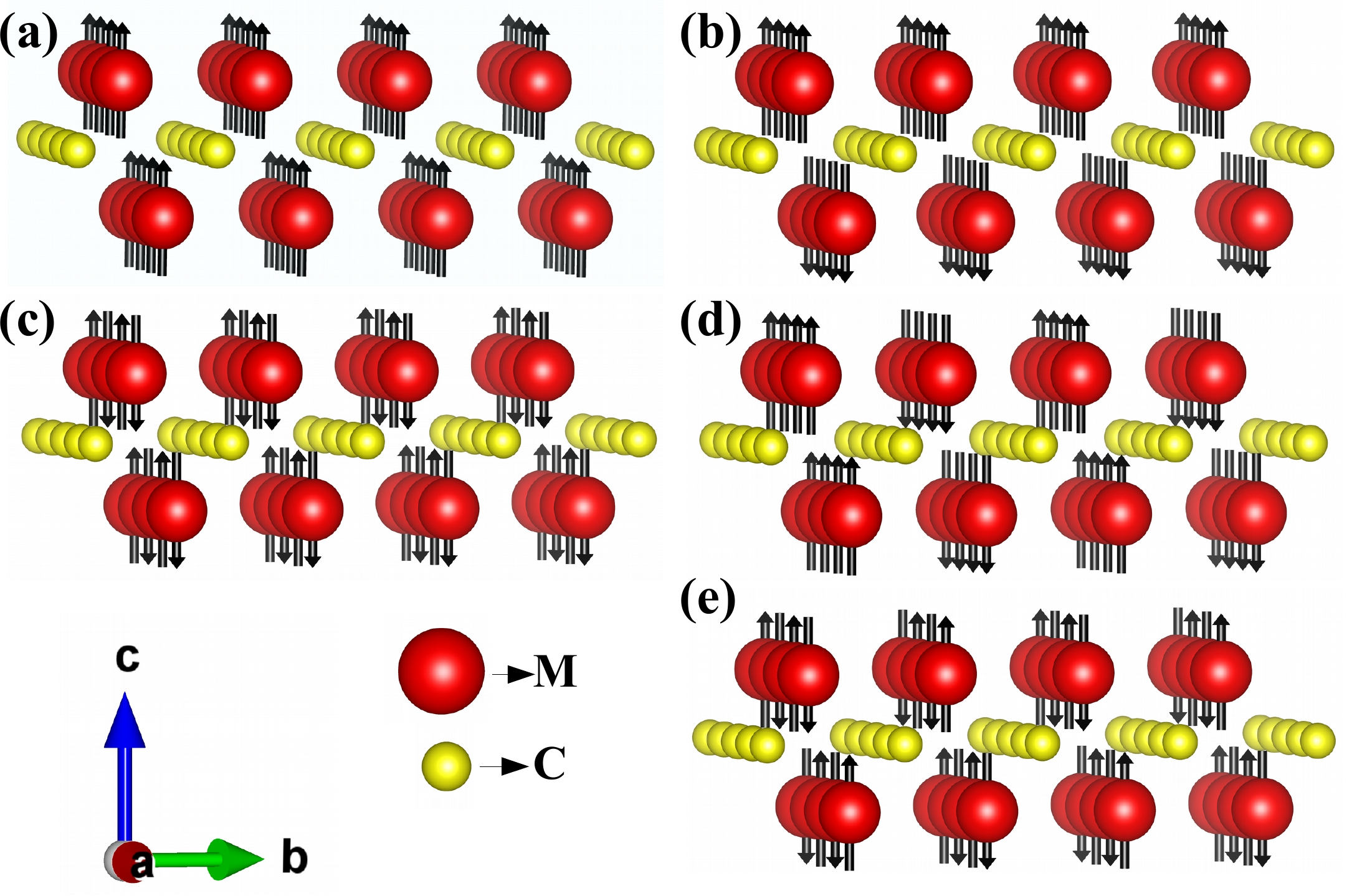}
    \caption{Illustrative representations of different ordered magnetic configurations considered in this work: (a)FM, (b)AFM-c, (c)AFM-a, (d)AFM-b, (e)AFM-abc .}
    \label{Fig:2}
\end{figure}

In order to investigate the role of magnetism on the functional properties, the first thing to determine is whether the ground states of the systems considered are magnetic. To do this, we first consider five different ordered spin configurations for spin-polarised calculations (Figure \ref{Fig:2}(a)-(e)), namely, A) Ferromagnetic where all M elements have their moments aligned along $c$-axis (Figure \ref{Fig:2}(a)), B) Anti-ferromagnetic-c, where M elements in the top(bottom) layers have their moments aligned(anti-aligned) along $c$-axis (Figure \ref{Fig:2}(b)) but in-plane moments align, C) Anti-ferromagnetic-a where the moments are anti-aligned along $a$-axis but are aligned along other two directions (Figure \ref{Fig:2}(c)), D) Anti-ferromagnetic-b where  (Figure \ref{Fig:2}(d)), the moments are anti-aligned along $b$-axis but are aligned along other two directions and E) Anti-ferromagnetic-abc where the system overall has an anti-ferromagnetic structure (Figure \ref{Fig:2}(e)). For a given chemical composition, the energetically lowest state obtained from total energy calculations on all five magnetic configurations is identified as the ground state of the system.
With the increase in temperature, the spin ordering gradually gives way to a partial and finally complete spin-disordered paramagnetic state. A key component in the present work is to investigate the impact of the magnetic disorder on the electrochemical properties of the MXenes considered. One may think of turning the  spin polarisation off to model the paramagnetic state, as the total magnetic moment is still zero in the paramagnetic state. This would, nevertheless, amount to the loss of information on magnetic interactions among transition metal atoms which are very much prevalent in the paramagnetic state. Moreover, the modeling of the partial magnetic disordered states that are indirect representations of different temperatures of the systems would not be possible to model. A solution to this problem is wonderfully addressed by the Disordered Local Moment (DLM) method \cite{dlm,dlm1}. In this method, each lattice site having magnetic constituent $M$ is populated with a 50-50 binary disordered alloy M$^{\uparrow}_{0.5}$M$^{\downarrow}_{0.5}$; the $\uparrow(\downarrow)$ indicate moments aligning along (against) the magnetization axis. 
\begin{figure}[ht!]
    \includegraphics[width=0.48\textwidth,center]{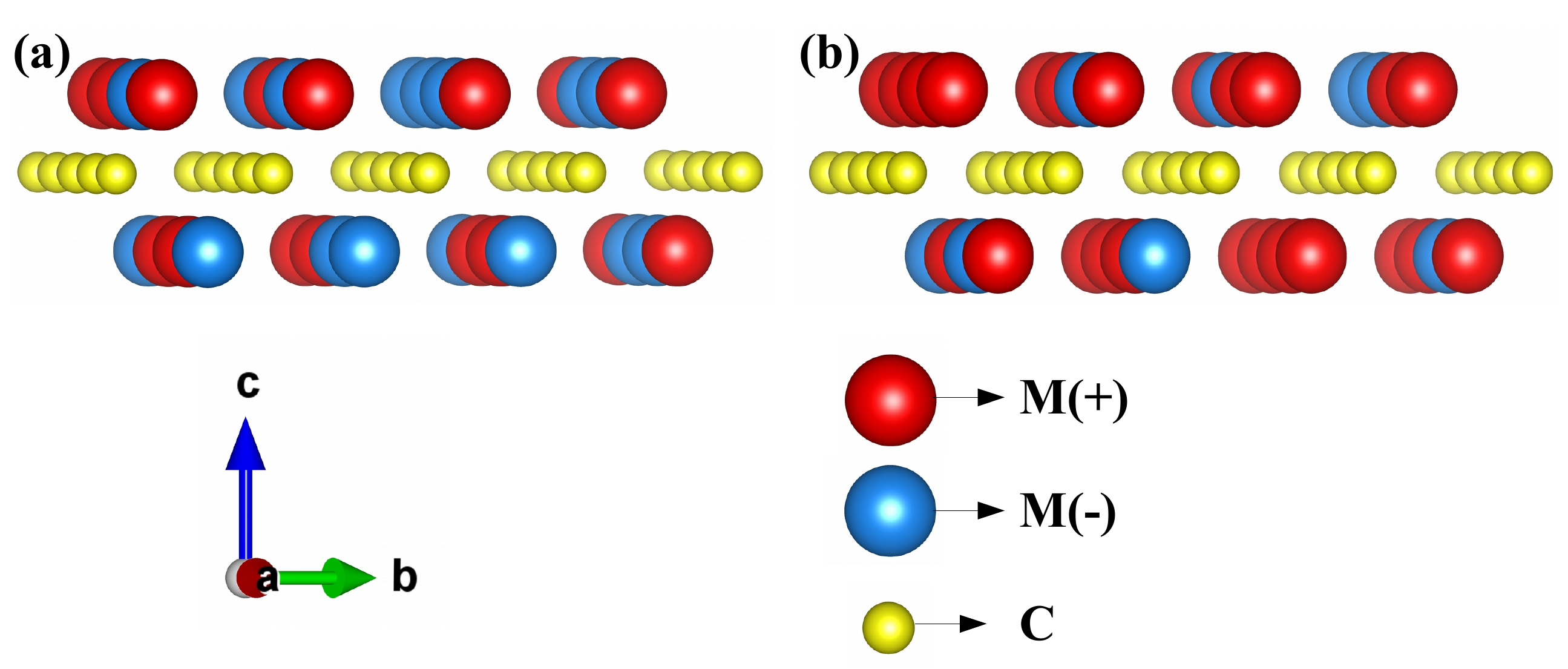}
    \caption{Illustrative representation of the arrangement of spin moments under the DLM picture used in this work: (a) a DLM system M$^{\uparrow}_{0.5}$M$^{\downarrow}_{0.5}$, and (b) a PDLM M$^{\uparrow}_{0.75}$M$^{\downarrow}_{0.25}$. M$^{\uparrow(\downarrow)}$ atoms are shown by red(blue) balls.The yellow balls represent the C atoms.}
    \label{Fig:3}
\end{figure}
This model thus simulates the paramagnetic state where local moments exist, but due to the disorder, the global magnetization vanishes. Consequently, a partial disordered local moment (PDLM) state that models partially ordered magnetic structure is obtained by considering a binary alloy M$^{\uparrow}_{1-y}$M$^{\downarrow}_{y}$ at the site of the magnetic constituent $M$ where $0.5<y<1$. In a supercell setup, we obtain a DLM state by randomly orienting the moments of atoms in the direction of the magnetization axis, keeping the number of sites where moments are aligning with the magnetization axis equal to the number of sites where moments are anti-aligning (Figure \ref{Fig:3} (a)). The PDLM state M$^{\uparrow}_{1-y}$M$^{\downarrow}_{y}$ is obtained by randomly picking up sites $N$, out of total $M$ sites, where the spin configuration of $N((M-N))$ sites are $\uparrow(\downarrow)$. A PDLM system with $y=0.75$ is schematically shown in Figure \ref{Fig:3}(b). In this work, for each of the three chemically disordered V$_{2-x}$Mn$_x$CO$_{2}$ MXenes, identified by their $x$ values, we have considered several PDLM states by changing the ratio $\frac{y}{1-y}$ of moments aligning with the magnetization axis and moments aligning against it. Each PDLM state is identified by a parameter $\eta = \frac{n^{\uparrow}-n^{\downarrow}}{n^{\uparrow}+n^{\downarrow}}$, implying the degree of magnetic disorder; $n^{\uparrow}(n^{\downarrow})$ are the number of transition metal sites where spin moments aligned along (against) the magnetization direction. A decrease in $\eta$, therefore, corresponds to an increase in magnetic disorder in the system. Each PDLM state, thus obtained, is stabilized by performing a series of DFT+U calculations \cite{dftu} where the Hubbard $U$ parameters for Mn and V are systematically reduced from a high value to zero. This procedure not only stabilizes the PDLM states but also stabilizes the individual magnetic moments of Mn and V, converging them to values close to their ground states for a given solid solution system. 
\subsection{The Electrochemical parameter: Total Capacitance with various contributions \label{capacitance}}
The measure of the electrochemical performance of an electrode material is its capacity to store charge. The appropriate quantity, therefore, is its capacitance when the material is in contact with an electrolyte.
The total capacitance ($C_T$) of an electrochemical capacitor is given as 
\begin{equation}
    \frac{1}{C_{T}} = \frac{1}{C_{Q}} + \frac{1}{C_{E}}
    \label{EQN: 2}
\end{equation}
$C_Q$ and $C_E$ are the quantum and electrical capacitances of the electrode, respectively. The \textquote{Quantum Capacitance} C$_Q$ depends on the electronic structure of the electrode material around the Fermi level. Unlike bulk, $C_Q$ becomes non-negligible in two dimensions \cite{luryi1988quantum} as in the low-dimension $C_Q$ values are of the same order as $C_E$. $C_Q$ is calculated from the electronic densities of states(DOS) of the material under consideration
\begin{equation}
   C^{diff}_{Q} = \frac{dQ}{dV} = e^{2}\frac{dN}{dE} =  e^{2}\int^{+\infty}_{-\infty} D(\xi)F_{T}(\xi-E)d\xi 
\label{EQN: 3}
\end{equation}
\begin{equation}
         C^{int}_{Q} = \frac{1}{eV}\int^{V}_{0} C^{diff}_{Q} dV
         \label{EQN: 4}
\end{equation}
$C^{diff}_{Q}$ and $C^{int}_Q$ are the differential and integrated quantum capacitances, respectively. $D(\xi)$, $F_{T} \left(\xi - E \right)$, and $E$ are the densities of states, thermal broadening function, and energy corresponding to the applied voltage, respectively.

An electrochemical capacitor stores charge through two mechanisms: the Electrochemical Double Layer effect (EDL) quantified by $C_{EDL}$ and the Redox mechanism quantified by $C_{redox}$. The electrical capacitance of the electrode is the resultant of the parallel combination of $C_{EDL}$ and $C_{redox}$
\begin{equation}
    C_E  =  C_{EDL} + C_{redox}
    \label{EQN: 5}
\end{equation}
We calculate the $C_{redox}$ using the formalism by Wang $et. al.$\cite{wang2019origin}. This formalism is based upon the following considerations: a) the electrode is negatively charged, b) the electrolyte contains H$^+$ ion. The electrode-electrolyte interaction enables the H$^+$ ion adsorption on the electrode surface. The systematic adsorption of H$^{+}$ results in a rigid shift of the electronic structure, the so-called \textquote{Rigid Band Approximation} (RBA). In this formalism,  $C_{redox}$ is computed as
\begin{equation}
    C_{redox} = \frac{\Delta Q}{\Delta V}
    \label{EQN: 6}
\end{equation}
The charge transfer between the electrode and H$^{+}$-ion across the double-layer and the potential change due to this charge transfer is given by $\Delta Q$ and $\Delta V$, respectively. $\Delta V$ is related to the change in the work-function ($\Delta_{WF}$) and  is given by
\begin{equation}
    \Delta WF = WF - (WF^{neutral}-\Delta E_{F})
    \label{EQN: 7}
\end{equation}
$WF$, $WF^{neutral}$, and $\Delta E_F$ are the work function of the H-adsorbed electrode, the work function of the neutral electrode, and the change in Fermi-level due to H adsorption. The work function of the electrode is evaluated from the energy of the vacuum ($E_{vac}$) and the Fermi-energy(E$_F$).
\begin{equation}
    WF = E_{vac}-E_{F}
    \label{EQN: 8}
\end{equation}
$C_{EDL}$ is evaluated the following way
\begin{equation}
    C_{EDL} = \frac{\epsilon_{0}\epsilon_{r}A}{d},
    \label{EQN: 9}
\end{equation}
$\epsilon_{0}$, $\epsilon_{r}$, $A$ and $d$ are the permittivity of the free space, the dielectric constant of the electrolyte, area of the electrode material, and width of the double-layer layer formed at the interface. 
\subsection{Computational Details}
The electronic structure calculations are performed by the DFT-based projector augmented wave method \cite{paw}as implemented in Vienna ab-initio Simulation Package(VASP)\cite{kresse1999ultrasoft}. The exchange-correlation part of the Hamiltonian is described by the Generalized Gradient Approximation (GGA)\cite{perdew1996generalized}. We use a kinetic energy cut-off of 520 eV and a Monkhrost-pack \cite{mp} grid of 4$\times$4$\times$1 for self-consistent calculations. A larger k-mesh of 6$\times$6$\times$1 is used for the  Density of States calculations. The energy and force criteria were set to 10$^{-6}$ eV and 0.05 eV/$\AA$.

\section{Results and Discussions}
\subsection{Ground state properties\label{result-structure}}
\begin{figure}[ht!]
    \includegraphics[width=1.0\linewidth,center]{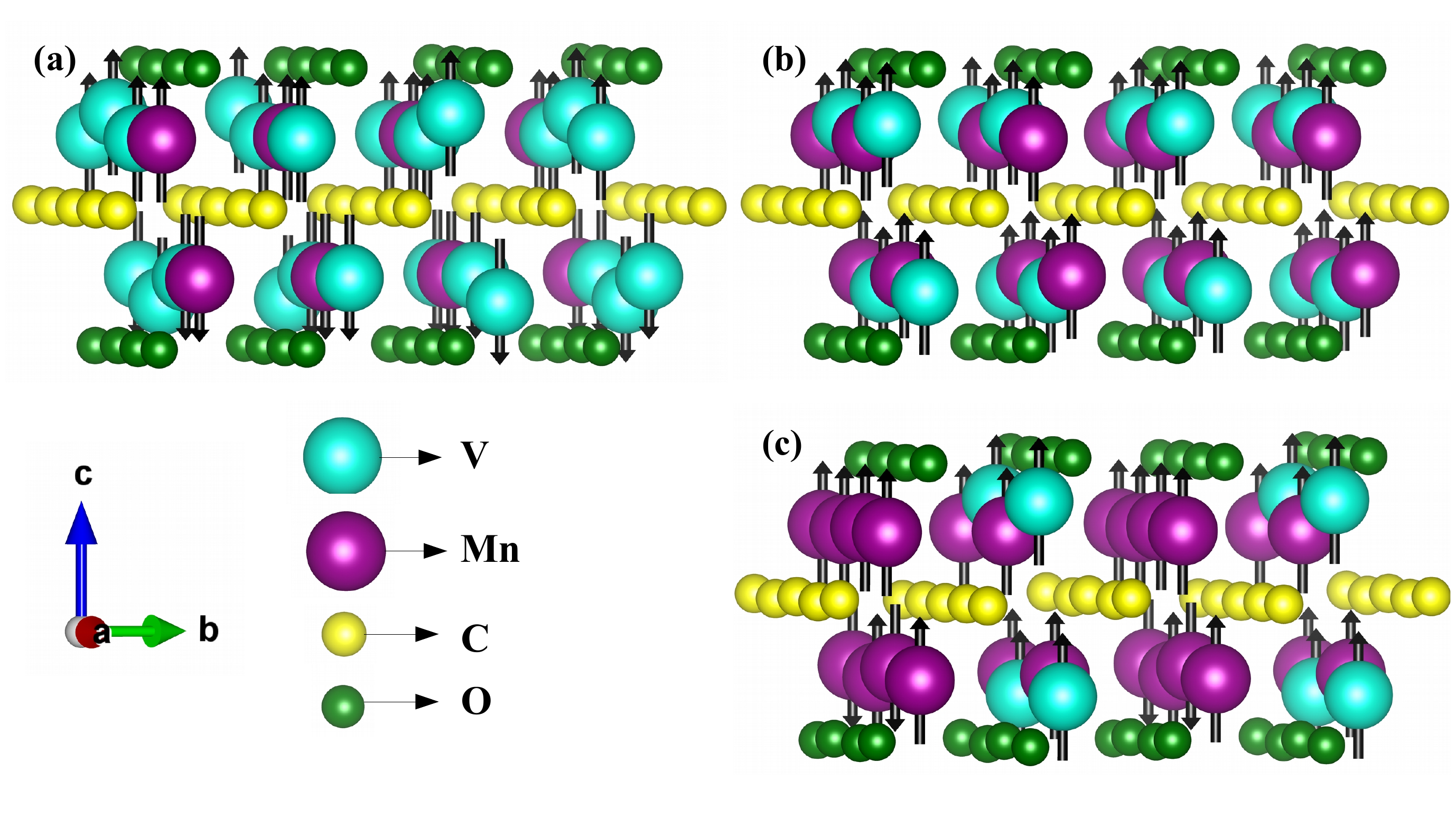}
    \caption{Magnetic ground state of A. V$_{1.5}$Mn$_{0.5}$CO$_2$, B. V$_{1.0}$Mn$_{1.0}$CO$_2$ and C. V$_{0.5}$Mn$_{1.5}$CO$_2$}
    \label{Fig:4}
\end{figure}
\begingroup
\begin{table*}[ht!]
\setlength{\tabcolsep}{8pt} 
\renewcommand{\arraystretch}{1.0} 
    \caption{Lattice-parameters($a$), magnetic ground states, and magnetic moments of constituent atoms of V$_{2-x}$Mn$_x$CO$_2$}
    \begin{tabular}{|c@{\hspace{0.2cm}}| c@{\hspace{0.2cm}}| c@{\hspace{0.3cm}}| cccc@{\hspace{0.5cm}}|} 
    \hline
    \vspace{-0.33cm}
    \\ System & $a$($\AA$)  &  magnetic ground state & 
    \multicolumn{4}{c|}{m($\mu_B$)}\\
    \hline
               &    &    &    Mn &  V  &  C  &  O   \\  
    \hline
    V$_{1.5}$Mn$_{0.5}$CO$_{2}$ & 11.64 & AFM-c &  +2.09 , -2.09 & 0.02 & -0.07 & -0.01\\
    V$_{1.0}$Mn$_{1.0}$CO$_{2}$ & 11.64 & FM &  2.422 & 0.053 & -0.03 & -0.01 \\
    V$_{0.5}$Mn$_{1.5}$CO$_{2}$  &  11.48  & FiM  &  2.7, 2.6,-1.7 & 0.38, 0.14 & -0.07 & -0.02 \\
    \hline
    \end{tabular}
    \label{Tab:1}
\end{table*}    
\endgroup
The ground states of V$_{2-x}$Mn$_{x}$CO$_{2}$ for three different $x$ are obtained the following way: non-spin-polarised calculations are first done to find the optimized structure of V$_{2-x}$Mn$_{x}$C. For each of the three solid solution compounds, six different calculations by placing the oxygen atoms at different positions (A, B, T as shown in Figure \ref{Fig:1}) on both surfaces are then done to determine the preferable sites of -O functionalization. After optimization of the functionalized solid solution MXenes spin polarised calculations are done for each of the five magnetic configurations (Figure \ref{Fig:2}). The lowest energy structure obtained from the calculations is considered the final ground state of V$_{2-x}$Mn$_{x}$CO$_{2}$ for a given $x$. The ground states for the three compounds are shown in Figure \ref{Fig:4}, and the relevant properties are listed in Table \ref{Tab:1}.   
Figure \ref{Fig:4} shows the ground state structures for the three MXene solid solutions. In the ground states of the compounds, the favorable positions of the oxygen atoms are over the hollow sites of the transition metal atoms ( Position \textquote{A} in Figure \ref{Fig:1}). This is consistent with the fact that for the end-point compounds Mn$_2$C and V$_2$C, this is the preferred site for the oxygen functional group. However, the surfaces of the solid solution MXenes are rumpled. This rumpling is caused by significant differences in the distances between the two transition metal atom types on the top and bottom surfaces. This, once again, follows the trend found in the end-point compounds. The distance between M$_1$ and M$_2$ layers in V$_2$C (Mn$_2$C) is 2.18{$\AA$}(2.0{$\AA$}).Such differences are observed even in the solid solution of MnVCO$_{2}$. The V$_1$-V$_2$ (Mn$_1$-Mn$_2$) distances are 3.00{$\AA$},2.68{$\AA$} and 3.06{$\AA$} (2.22{$\AA$},2.2{$\AA$} and 2.21{$\AA$}) for $x=0.5,1.0$ and $1.5$, respectively.

From Figure \ref{Fig:4} and Table \ref{Tab:1}, it is evident that the magnetic ground states change with changes in the chemical composition. While the ground state of V$_{1.5}$Mn$_{0.5}$CO$_{2}$ is AFM-c, we get a ferromagnetic (FM) ground state for V$_{1.0}$Mn$_{1.0}$CO$_{2}$. The magnetic ground state of V$_{0.5}$Mn$_{1.5}$CO$_{2}$ is quite unique. For this compound in the ground state, the top surface comprises Mn spins aligned along $c$-direction with moments 2.7 and 2.6 $\mu_{B}$. The bottom surface, on the other hand, has a Ferrimagnetic alignment of Mn spins. On this surface, one-third of the Mn atoms align anti-parallel to the other Mn atoms. Each one of these anti-aligned Mn is surrounded by four aligned Mn atoms in a hexagonal environment while each Mn atom, aligned along $c$, is in the surrounding environment of four Mn atoms, out of which two are also aligned. Two V atoms occupy the remaining vertices of this hexagonal network. The arrangement of atoms on the surfaces of this compound is shown in Figure \ref{fig:1}(a) and \ref{fig:1}(b) of Appendix A. In bottom surface the atoms Mn$_{13}$-Mn$_{16}$ (Figure \ref{fig:1}(b)) are anti-aligned whereas the rest are aligned. If one considers the hexagon with Mn$_{16}$, an atom whose spin is aligned opposite to $c$-direction (the other three with the same spin direction, out of total 12 atoms on this surface, are Mn$_{13}$-Mn$_{15}$), at the center, the four Mn atoms Mn$_{19}$, Mn$_{20}$, Mn$_{22}$and Mn$_{24}$ occupying vertices of the hexagon, have their moments aligned along $c$. On the other hand, if the hexagon with Mn$_{19}$ at the center is considered, Mn$_{21}$ and Mn$_{24}$ having aligned moments occupy two vertices while Mn$_{15}$ and Mn$_{16}$, the two having spins anti-aligned, occupy other two vertices. Such arrangements make the magnetic environment around Mn atoms in the bottom surface inhomogeneous though the chemical environment is homogeneous. The magnetic moments on anti-aligned Mn atoms are significantly reduced to 1.7 $\mu_{B}$ while the aligned Mn atoms have a moment of 2.7 or 2.6 $\mu_{B}$ on them. Upon inspecting the Mn-Mn bond distances, we find the following: (a) The Mn atoms in the top layer having a moment of 2.7 $\mu_{B}$ are connected to four Mn atoms sitting on the vertices of the hexagon around them with Mn-Mn bond distances along a line connecting three Mn atoms ( which means two Mn-Mn bonds) between 2.84-2.86 $\AA$ and 2.92-2.94 $\AA$ (For example, consider Mn$_{4}$ in Figure \ref{fig:1}(a). The Mn$_{8}$-Mn$_{4}$, Mn$_{7}$-Mn$_{4}$,Mn$_{10}$-Mn$_{4}$ and Mn$_{12}$-Mn$_{4}$ bond distances are 2.83 $\AA$, 2.93 $\AA$, 2.86 $\AA$ and 2.92 $\AA$, respectively). The ones that have their moments slightly reduced have one of the bonds shrunk to 2.75 $\AA$  while the other along the same line joining the three Mn elongated to 3.02 $\AA$ (For example, Mn$_{8}$-Mn$_{10}$ and Mn$_{5}$-Mn$_{10}$ bond lengths are 2.75 $\AA$ and 3.02 $\AA$ respectively, while Mn$_{2}$-Mn$_{10}$ and Mn$_{4}$-Mn$_{10}$ distances are more uniform, 2.86 $\AA$ and 2.92 $\AA$, respectively). The Mn-V bond distances vary too; for Mn$_{1}$-Mn$_{4}$, the bond distances are 2.89 $\AA$ and 2.93 $\AA$ while for the rest 8 Mn, they are 2.84 $\AA$ and 3.0 $\AA$ rendering the bond lengths along a V-Mn-V line non-uniform, (b) in the bottom layer, when an anti-aligned Mn is connected to four aligned Mn, the two bond distances along a line connecting three Mn atoms are significantly non-uniform, one between 2.68 and 2.72 $\AA$ and the other between 3.06 and 3.1 $\AA$ (For example, in Figure \ref{fig:1}(b), Mn$_{16}$-Mn$_{22}$, Mn$_{16}$-Mn$_{24}$, Mn$_{16}$-Mn$_{19}$ and Mn$_{16}$-Mn$_{20}$ bond lengths are 3.06 $\AA$, 2.72 $\AA$, 2.68 $\AA$, and 3.1 $\AA$, respectively). When an aligned Mn is connected to two anti-aligned atoms, the distribution of bond distances are same. However, the two bond distances for bonds connecting three aligned Mn atoms are 2.77 $\AA$ and 3 $\AA$ (For example, Mn$_{19}$-Mn$_{24}$ and Mn$_{19}$-Mn$_{21}$ shown in Figure \ref{fig:1}(b)), similar to the case of Mn in the top layer having a moment of 2.6 $\mu_{B}$. The moments of all Mn atoms for compositions with $x=0.5,1.0$ are the same; the Mn-Mn bond distances are uniform too. The atomic arrangements on the surfaces of V$_{1.0}$Mn$_{1.0}$CO$_{2}$ and V$_{1.5}$Mn$_{0.5}$CO$_{2}$ are shown in Figures \ref{fig:1}(c-d) and \ref{fig:1}(e-f) of Appendix A, respectively. For these two compounds, the Mn moment increases with an increase in Mn concentration, an expected outcome due to the presence of more Mn in the neighborhood of each Mn. One more noteworthy result is the different V moments of the atom on the two surfaces of $V_{0.5}$Mn$_{1.5}$CO$_{2}$. While the V atoms in the bottom layer, located at the center of a hexagon whose vertices have Mn atoms only, have moments of only 0.14 $\mu_{B}$, the ones in the top layer, sitting in the same environment of Mn, have moments of 0.38 $\mu_{B}$ localized on them. Here too we find an asymmetry in the Mn-V bond distances in the two surfaces. While the five Mn-V bond distances of the bottom surface vary between 2.87-2.94 $\AA$, one of the Mn-V bonds in the top surface is elongated to 3 $\AA$. Since the spin polarisation in V must be due to the magnetic Mn atoms, a slightly higher moment on V atoms in the top layer can be correlated with the non-uniform Mn-V bond distances. For the other two compounds, there is hardly any induced moment on V. The results suggest that the magnetic configurations on the surface and the bond lengths between transition metal atoms are correlated. This is likely to influence the localization of charges along a chemical bond. This, in turn, should have impacts on the surface-related phenomena, the main interest of this paper. 

\subsection{Electronic Structure: effect of chemical and magnetic disorder\label{resutl-electronic}}
\begin{figure}
    \centering
    \includegraphics[width=1.0\linewidth]{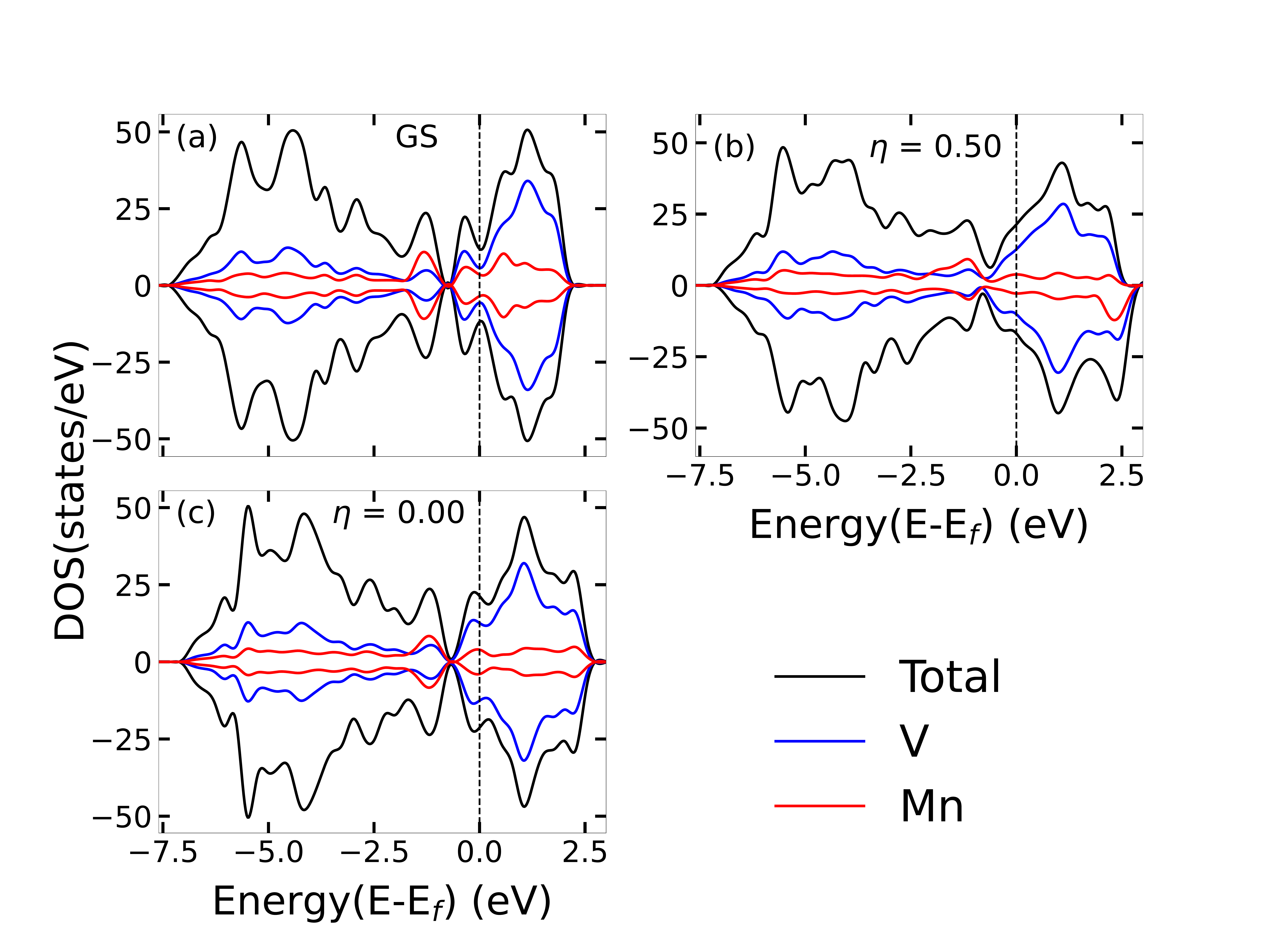}
    \caption{Local Densities of States of V$_{1.5}$Mn$_{0.5}$CO$_2$ in (a)Ground , (b)$\eta$=0.5 and (c)$\eta$=0.0 state.}
    \label{Fig:5}
\end{figure}

\begin{figure}
    \centering
    \includegraphics[width=1.0\linewidth]{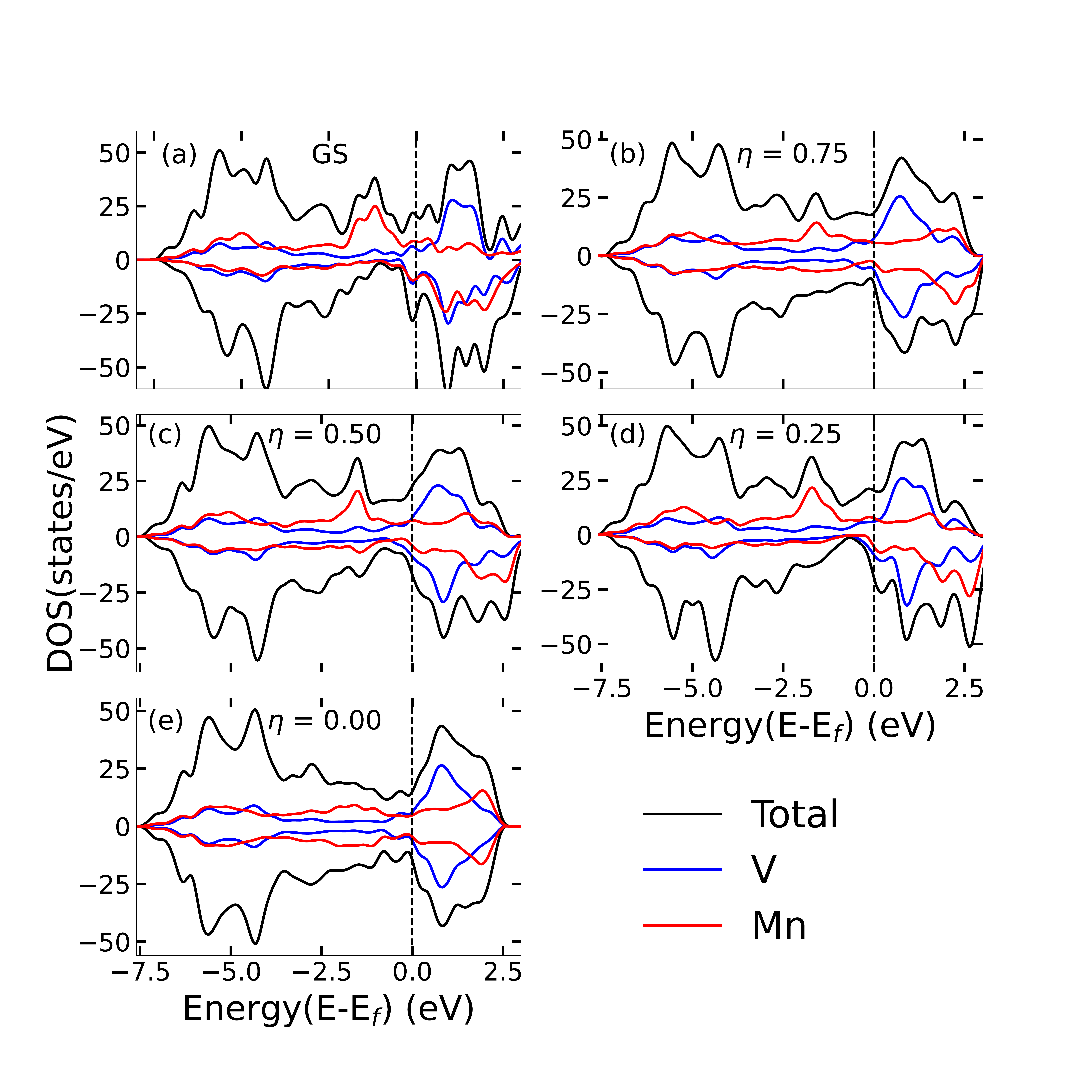}
    \caption{Local Densities of States of V$_{1.0}$Mn$_{1.0}$CO$_2$ in (a)Ground , (b)$\eta$=0.75, (c)$\eta$=0.50, (d)$\eta$=0.25 and (e)$\eta$=0.00 state.}
    \label{Fig:6}
\end{figure}

\begin{figure}
    \centering
    \includegraphics[width=1.0\linewidth]{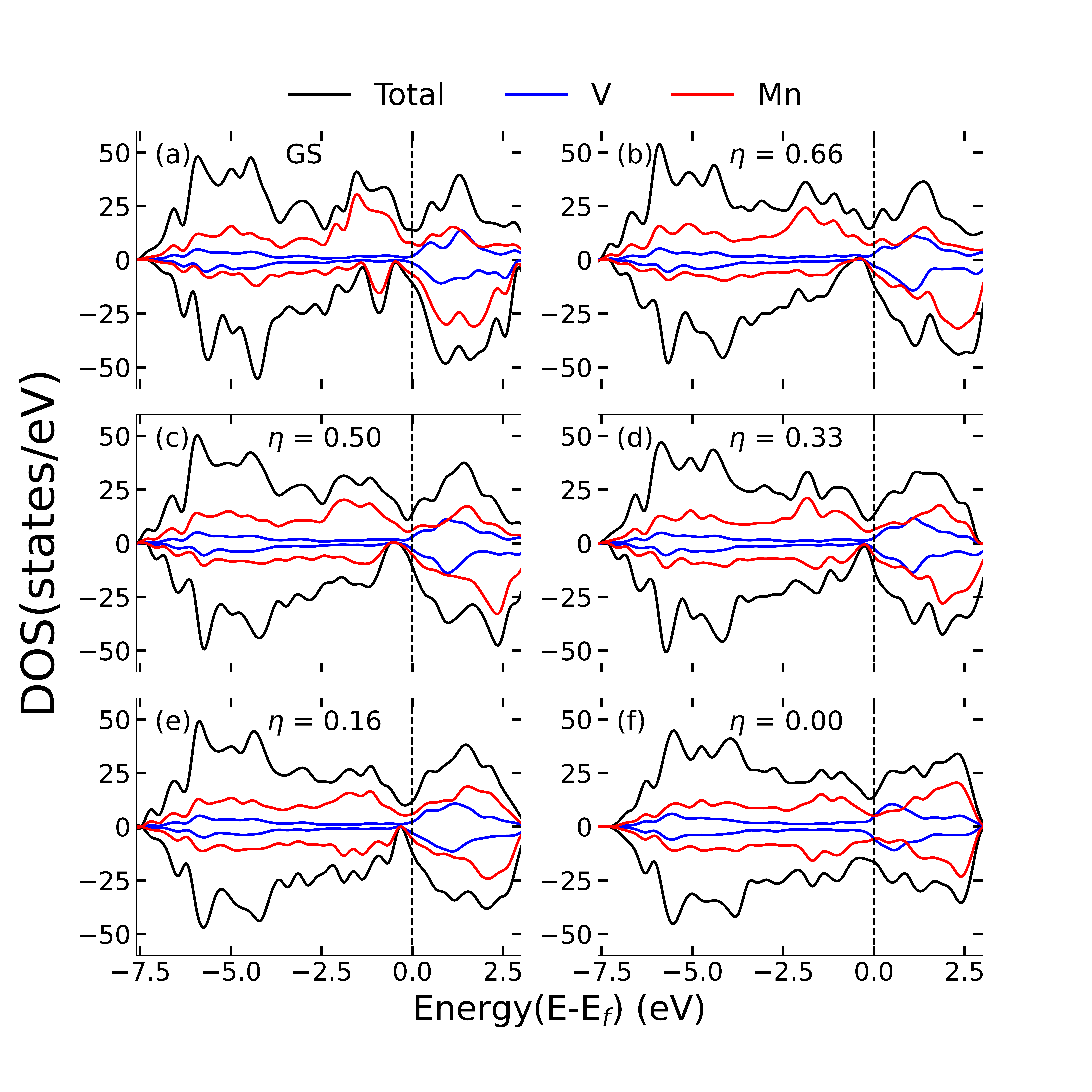}
    \caption{Local Densities of States of V$_{0.5}$Mn$_{1.5}$CO$_2$ in (a)Ground , (b)$\eta$=0.66, (c)$\eta$=0.50, (d)$\eta$=0.33, (e)$\eta$=0.16, and (f)$\eta$=0.00 state.}
    \label{Fig:7}
\end{figure}

The changes in the electronic structures due to changes in chemical composition and magnetic order are expected to be significant and throw insights into the role of each type of disorder. This may have a substantial impact on the electrochemical properties. Quantum capacitance is directly related to the electronic structure of the material; the changes in the charge transfer due to disorder will have effects on the electrical capacitance. Therefore, in Figures (\ref{Fig:5}-\ref{Fig:7}), we have shown the total and atom projected densities of states of V$_{2-x}$Mn$_{x}$CO$_{2}$ for different chemically and magnetically disordered compounds. Since Mn and V atoms should play key roles, only their partial densities of states are shown. For each $x$ that is a solid solution, we investigate the changes in the densities of states as $\eta$, the degree of magnetic disorder changes. We first analyze the ground states of the three compounds, then the changes from the ground state to a particular partial magnetic disordered state (a state with the same $\eta$) for each $x$, and finally compare the electronic structures of full magnetic disordered states for different $x$.  

Though the ground state magnetic order is different in the three compounds, there are a few noticeable features that follow a trend. There is a sharp pseudo-gap around -1 eV in both majority and minority spin bands of V$_{1.5}$Mn$_{0.5}$CO$_2$. The symmetry with respect to the spin band is due to AFM-c magnetic order in this compound. This pseudo-gap turns shallow and nearly non-existent in V$_{1.0}$Mn$_{1.0}$CO$_2$. With further increase in Mn content, the pseudo-gap somewhat re-appears in the ground state of V$_{0.5}$Mn$_{1.5}$CO$_2$ at an energy closer to the Fermi energy and in the minority spin band only. The partial densities of states for the three compounds clearly show that the V bands are nearly unoccupied irrespective of the composition and the magnetic ground state. This explains the near-zero or at the most, a small V moment. In order to understand the electronic structures better, we have shown the partial densities of states of all Mn (V) atoms in the supercell for the three compounds in Figures \ref{fig:2}(\ref{fig:3}), \ref{fig:4}(\ref{fig:5}) and \ref{fig:6}(\ref{fig:7}) of Appendix A. The sharp pseudo-gap in the electronic structures of V$_{1.5}$Mn$_{0.5}$CO$_2$ is an artifact of strong hybridization of Mn $d$, V $d$, and O $p$ states.
In this compound, the electronic structure of each Mn atom is identical which is a reflection of an identical environment around them, in terms of chemical specie and the bond distances. Apart from Mn-V bond distances that are uniformly distributed with the hexagon around an Mn atom, Mn-O, and Mn-C bonds are nearly uniform (Mn-O bond lengths vary between 1.96 $\AA$ and 2.05 $\AA$ while Mn-C bond lengths vary between 2.03 $\AA$ and 2.08 $\AA$). The V partial densities of states, on the other hand, can clearly be grouped into two bunches. Eight V atoms (V$_{2}$, V$_{4}$, V$_{9}$, V$_{11}$, V$_{14}$, V$_{16}$, V$_{21}$ and V$_{23}$, Figure \ref{fig:3}, Appendix A) have identical densities of states with a pseudo-gap like feature around -1 eV in both majority and minority bands. For these V atoms, out of the three V-O bonds, two are significantly smaller (1.76 $\AA$) in comparison to the other (1.88 $\AA$). For the other 16 V atoms, V-O bonds are much longer, between 1.93 and 2.12 $\AA$. Such shorter V-O bonds result in stronger hybridization between $d$ orbitals of V and $p$ orbitals of O, leading to the pseudo-gap in the electronic structures. This pseudo-gap-like feature is observed in the other 16 atoms at deeper energy. All Mn and V atoms in the ground state of V$_{1.0}$Mn$_{1.0}$CO$_{2}$ have identical electronic structures. This is expected as the chemical and magnetic environment around each atom is the same. The surfaces have identical spin alignments too. The effect of the increase in Mn content is visible if one compares it with the Mn-deficient compound. An increase in chemical disorder smears the main Mn peak observed in the Mn-deficient compound, producing a peak and a shoulder. The Mn-V bonds relax as compared to the ones in the Mn-deficient system. In V$_{1.5}$Mn$_{0.5}$CO$_{2}$, Mn-V bond distances were 2.72 $\AA$ and 3.07 $\AA$. In V$_{1.0}$Mn$_{1.0}$CO$_{2}$, they are 2.82 $\AA$, 2.86 $\AA$, 2.99 $\AA$ and 3.02 $\AA$. Such relaxations in the bond lengths reduce the Mn-V hybridizations resulting in the pseudo-gap completely vanishing in the majority spin band and becoming shallow in the minority one. Larger disorder in Mn-rich V$_{0.5}$Mn$_{1.5}$CO$_{2}$, as compared to Mn-deficient system now further delocalizes the Mn states (Figure \ref{fig:6}, Appendix A). However, sharp peaks and more localization of the majority states are observed for 4 Mn atoms (Mn$_{1}$-Mn$_{4}$) in the upper surface. For these Mn, the Mn-transition metal bond lengths are more uniformly distributed, varying between 2.83 $\AA$ and 2.99 $\AA$. For the other Mn atoms, two Mn bonds along a line joining 3 Mn atoms are severely non-uniform, 2.75 $\AA$ and 3.01 $\AA$. The Mn-V bond lengths corresponding to these Mn atoms follow the same trend. Thus, in spite of having identical chemical environments, the larger non-uniformity in bond distances leads to a slightly smaller exchange splitting for these Mn atoms and delocalization in Mn states. The electronic structures of the V atoms are identical, as expected. In the bottom surface, the inhomogeneity in the magnetic environments around a transition metal atom affects their electronic structures. Mn$_{13}$-Mn$_{16}$ have their spins aligned along $-c$ direction and have more localized Mn states as compared to the other 8 Mn atoms in this surface, which have spin orientations opposite to them. They are surrounded by four Mn with spins aligned along $c$ and two V atoms. On the other hand, any one of the other eight atoms on the surface is surrounded by two Mn, which have the same spin alignment as the atom in question, and two others that have spin alignment opposite to it. This leads to a frustrated triangular network of Mn atoms resulting in a more delocalized electronic structure. The electronic structures of the four V atoms on this surface are identical as they are in identical chemical and magnetic environments. A pseudo-gap-like feature is observed in their minority bands, suggesting slightly stronger hybridization in comparison to that in V$_{1.0}$Mn$_{1.0}$CO$_{2}$. This can be due to slightly shorter and near uniform Mn-V bonds(bond lengths varying between 2.89-2.94 $\AA$). The pseudo-gap-like feature is observed in the top surface Mn minority bands. There too, out of four Mn-V bonds, three vary between 2.84-2.93 $\AA$ only.

In Figure \ref{Fig:5}(b) and Figure \ref{fig:8}, Appendix A, we show the densities states of a PDLM state $\eta=0.5$ for V$_{1.5}$Mn$_{0.5}$CO$_{2}$. The PDLM state is obtained by flipping the spins of one atom (Mn$_{4}$) from the top surface and three atoms from the bottom surface (Mn$_{6}$-Mn$_{8}$). Consequently, the loss of symmetry with respect to the spin orientation affects the electronic structures of different atoms in different ways. The Mn states, in general, are less localized. We find that the ground state-like features are preserved more for the pairs Mn$_{1}$-Mn$_{5}$ and Mn$_{4}$-Mn$_{8}$. Each pair of atoms are located along $c$-direction. Even after random flipping of spins, the spins of each pair are anti-aligned, mimicking the arrangement in the ground state. This is the reason behind the close resemblances in their electronic structures with that in the ground state. That the anti-alignment of spins along $c$-axis has a profound effect on the electronic structure becomes clear from the electronic structures for the complete magnetically disordered state $\eta=0$ (Figure \ref{Fig:5}(c) and Figure \ref{fig:11}, Appendix A. We find that the total densities of states for both spin bands are very similar to that of the ground state. The reason can be understood from the Mn densities of states (Figure \ref{fig:11}, Appendix A). One can see that the Mn densities of states in a complete spin-disordered state resemble closely those in the AFM-c ordered ground state. The reason lies in the anti-alignment of Mn spins along $c$-direction. After flipping more spins of $\eta=0.5$ state randomly to achieve a complete spin-disordered state on both surfaces, Mn$_{2}$-Mn$_{6}$ and Mn$_{3}$-Mn$_{7}$ pairs have anti-aligned arrangement of their spins along $c$-direction, along with the other two pairs that were anti-aligned in partial magnetic disordered state $\eta=0.5$. Such interactions along $c$-directions localized Mn states again, like the ground state.    

Total and Mn densities of states for $\eta=0.5$ state of V$_{1.0}$Mn$_{1.0}$CO$_{2}$ are shown in Figure \ref{Fig:6}(c) and Figure \ref{fig:9}, Appendix A. In comparison with the ground state, We find substantial changes in both spin bands, for energies near Fermi level. The peak in the minority band and the valley, flanked by two peaks, at the Fermi level of the ground state, are absent in this partially disordered state. Densities of states of different Mn atoms undergo noticeable changes from the magnetically ordered ground state. The presence of anti-aligned spins for pairs of Mn forces a change in the exchange field each Mn atom is subjected to resulting in a change in their electronic structures. Careful inspection of Mn densities of states reveals that Mn pairs comprising one Mn from the top surface and the other from the bottom, have identical densities of states that are different from one pair to the other. The pairs are Mn$_{1}$-Mn$_{16}$, Mn$_{2}$-Mn$_{11}$, Mn$_{3}$-Mn$_{10}$, Mn$_{4}$-Mn$_{9}$,Mn$_{5}$-Mn$_{15}$, Mn$_{6}$-Mn$_{14}$, Mn$_{7}$-Mn$_{13}$ and Mn$_{8}$-Mn$_{12}$. Mn spins in each pair align in the same direction. However, each pair has an Mn whose spin is anti-aligned to theirs in the neighborhood. The distances between the anti-aligned spins along with the details of the neighborhood make variations in the electronic structures of each pair. For example, each atom in the Mn$_{1}$-Mn$_{16}$ has an anti-aligned Mn atom at a distance of 2.91 $\AA$ while Mn$_{4}$-Mn$_{9}$ and Mn$_{5}$-Mn$_{15}$ pairs have the same at distances 5.84 $\AA$ and 6.5 $\AA$, respectively. When the magnetic disorder is complete ($\eta=0$), there are further changes in the densities of states close to the Fermi level (Figure \ref{Fig:6}(c). The symmetry is completely lost and each Mn has a distinctly different density of states (Figure \ref{fig:12}, Appendix A).    
 
In the case of $\eta=0.5$ partial disordered state of V$_{0.5}$Mn$_{1.5}$CO$_{2}$ too, the electronic structure near Fermi level changes considerably (Figure \ref{Fig:7}(c)). The pseudo-gap in the minority spin band is now wider in comparison with the ground state. The peak around -1 eV next to the pseudo-gap observed in the ground state has now disappeared. Upon inspection of the densities of states of each Mn atom (Figure \ref{fig:10}, Appendix A), we find that due to the inclusion of random flipping of Mn spins, the Mn states are more delocalized and that densities of states for any pair of Mn atoms are dissimilar. This is due to a complete lack of symmetry with respect to the spin structure. In the complete disordered phase ($\eta=0$), the electronic structure of each Mn undergoes further modifications (Figure \ref{fig:13}, Appendix A). Noticeably, the majority band in the electronic structures of those Mn atoms whose spins are flipped have filled the pseudo-gap. Complete randomness in the magnetic structure has thus weakened the hybridization between transition metal $d$ orbitals. 

The densities of states for various partially ordered magnetic states are shown in Figure \ref{Fig:6} and Figure \ref{Fig:7} for V$_{1.0}$Mn$_{1.0}$CO$_{2}$ and V$_{0.5}$Mn$_{1.5}$CO$_{2}$, respectively show that there are only slight changes near Fermi level in most cases when $\eta$ changes. It is also demonstrated that the ordering (or lack of it) of spins and not the number of spins with a particular orientation has a significant impact on the electronic structures. As an example, one can compare the ground state and $\eta=0.66$ state of V$_{0.5}$Mn$_{1.5}$CO$_{2}$. The differences in the total densities of states between the ordered ground state(Figure \ref{Fig:7}(a) and (b))and the partially ordered state are evident, even though both have the same number of spins aligned along $c$-direction. The Mn densities of states for this partially ordered magnetic state are shown in Figure \ref{fig:14}, Appendix A). We find that the random orientation of Mn spins in $\eta=0.66$ state has effected changes among the Mn atoms on the same surface and having the same spin orientation. This is an effect of the random exchange field produced by the neighborhood of the atoms. In a nutshell, the state of magnetic order has non-negligible effects on the electronic structures of V$_{2-x}$Mn$_{x}$CO$_{2}$ MXenes, particularly for the energies close to the Fermi level. Such changes are sure to affect the Quantum capacitance. In the next section, we present the results of this. 
\subsection{Effects of chemical and magnetic disorder on capacitances}
\begin{figure*}[ht!]
    \centering
    \begin{subfigure}[b]{0.3\textwidth}
    \includegraphics[width=\textwidth]{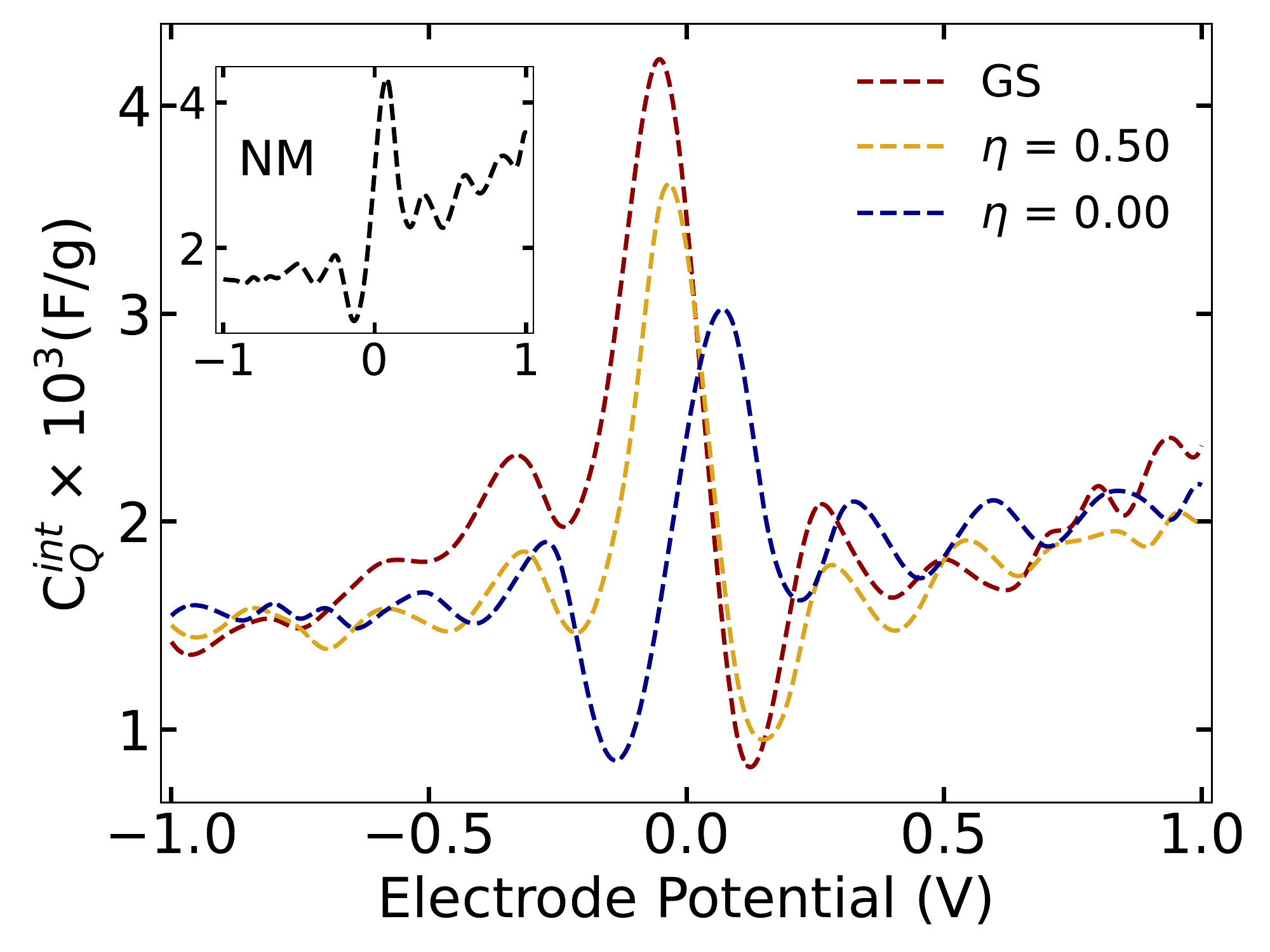}
    \caption{V$_{1.5}$Mn$_{0.5}$CO$_{2}$}
    \end{subfigure}
    \hspace{-0.05cm}
    \begin{subfigure}[b]{0.3\textwidth}
    \includegraphics[width=\textwidth]{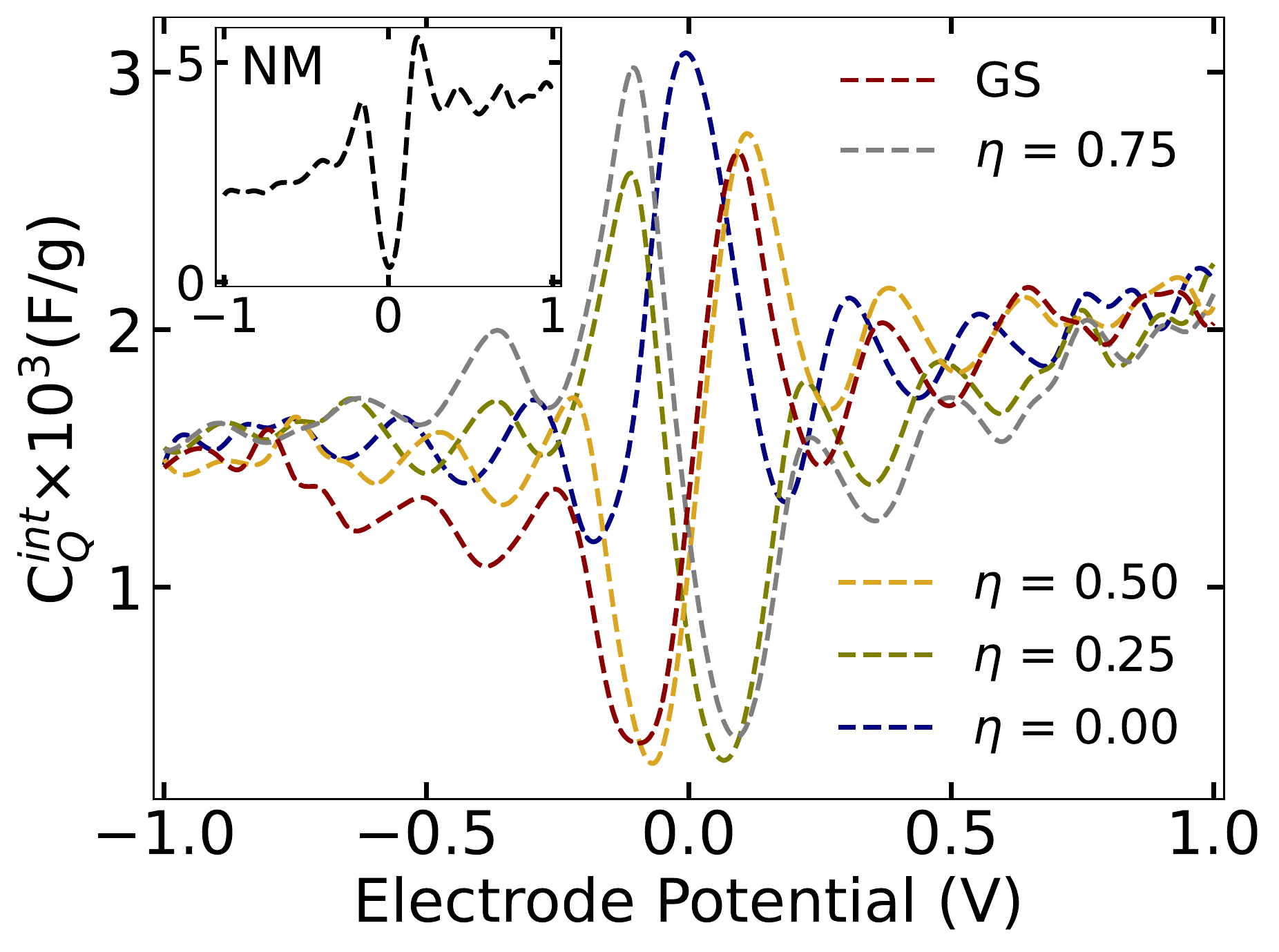} \caption{V$_{1.0}$Mn$_{1.0}$CO$_{2}$}
    \end{subfigure}
    \hspace{-0.05cm}
    \begin{subfigure}[b]{0.3\textwidth}
    \includegraphics[width=\textwidth]{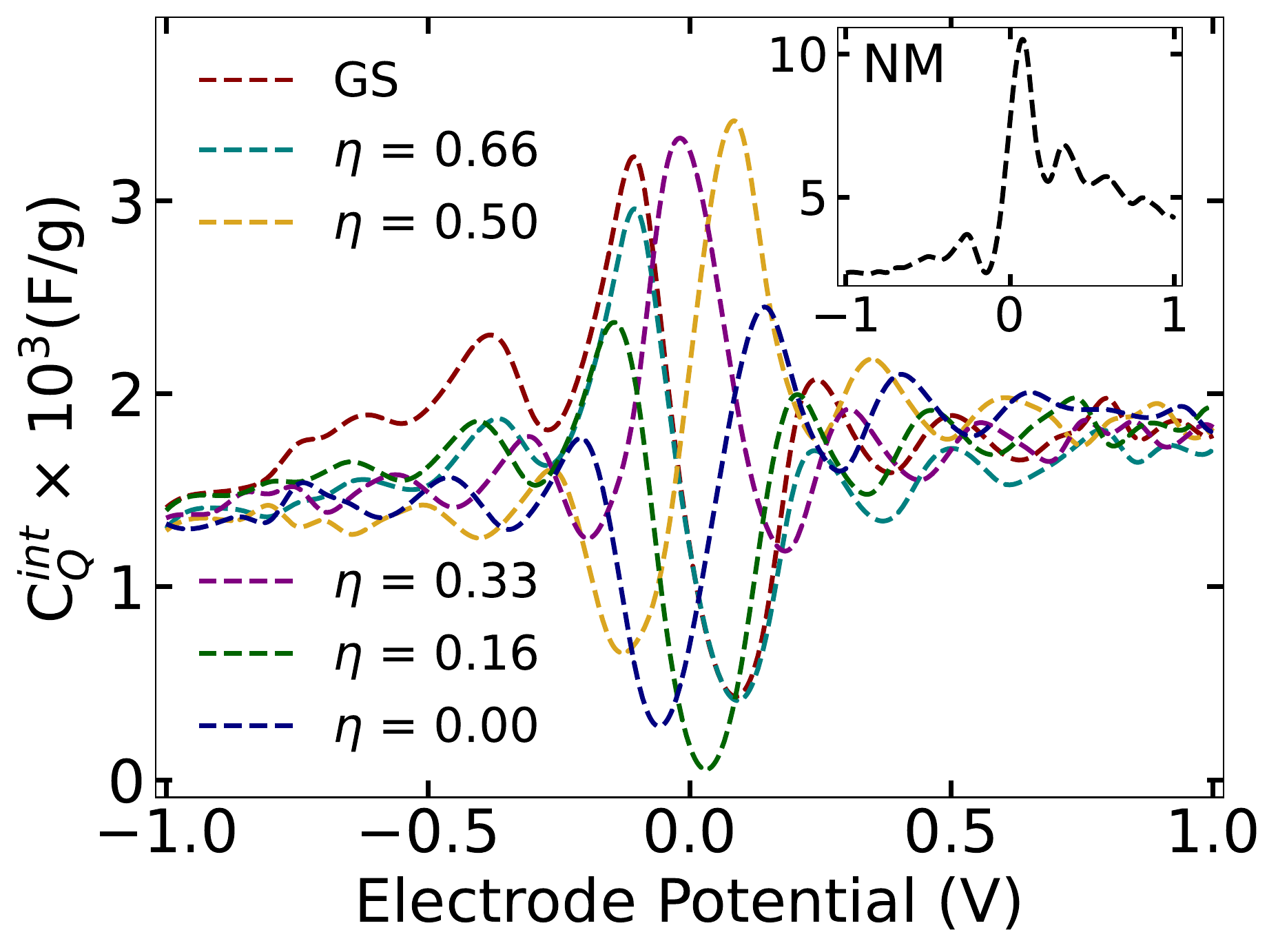}
    \caption{V$_{0.5}$Mn$_{1.5}$CO$_{2}$}
    \end{subfigure}
    \caption{Integrated Quantum Capacitance (C$^{int}_Q$) of ground state(GS) and different spin disordered states of V$_{2-x}$Mn$_{x}$CO$_{2}$(x=0.5,1.0,1.5). The non-magnetic behavior is shown in the inset of the respective plot. }
    \label{Fig:8}
\end{figure*}

In this section, we systematically discuss the changes in the quantum and electrical capacitance of V$_{2-x}$Mn$_{x}$CO$_{2}$ MXene electrodes when in contact with an acidic electrolyte like H$_{2}$SO$_{4}$. To this end, we first discuss the variations in the integrated quantum capacitance ($C_{Q}^{int}$) over a voltage window $\pm$1 V, for different $x$ and different magnetically ordered and disordered states. The results are shown in Figures \ref{Fig:8}(a)-(c).
$C_{Q}^{int}$ for non-magnetic states for each chemical composition are shown for the sake of comparison.$C_{Q}^{int}$ are calculated at room temperature. The wiggled structure in the $C_{Q}^{int}$ is an outcome of the modulations of the densities of states with the thermal broadening function (Equation\ref{EQN: 3},\ref{EQN: 4}). We find that irrespective of composition and magnetic state, the maximum of $C_{Q}^{int}$ occurs near 0V. This is commensurate with the fact that the changes in the electronic structures across the chemical and magnetic degrees of the disorder occur near the Fermi level. The maxima of quantum capacitance for non-magnetic states are found to be higher than the magnetic states, irrespective of the chemical composition. The maximum value increases with increasing $x$. However, this is due to the large densities of states at the Fermi level of non-magnetic states which signify instability of the non-magnetic phases. The variations in the maximum $C_{Q}^{int}$ for the magnetic ground state or any other magnetic disordered state, with $x$ are non-monotonic. This is due to the fact that the magnetic states, ground or disordered, are different for different $x$. However, since the typical voltage window used in the experiments are rather narrow and may not include 0V, it is more useful to look for the maximum $C_{Q}^{int}$ in the positive and negative voltage windows. Upon inspecting the variations in $C_{Q}^{int}$ in two windows separately, we do not find any specific trend except that the maximum value of $C_{Q}^{int}$ is 1500(2000) F/g in the negative(positive) voltage window. 
\begin{figure}[ht!]
    \includegraphics[width=0.7\linewidth,center]{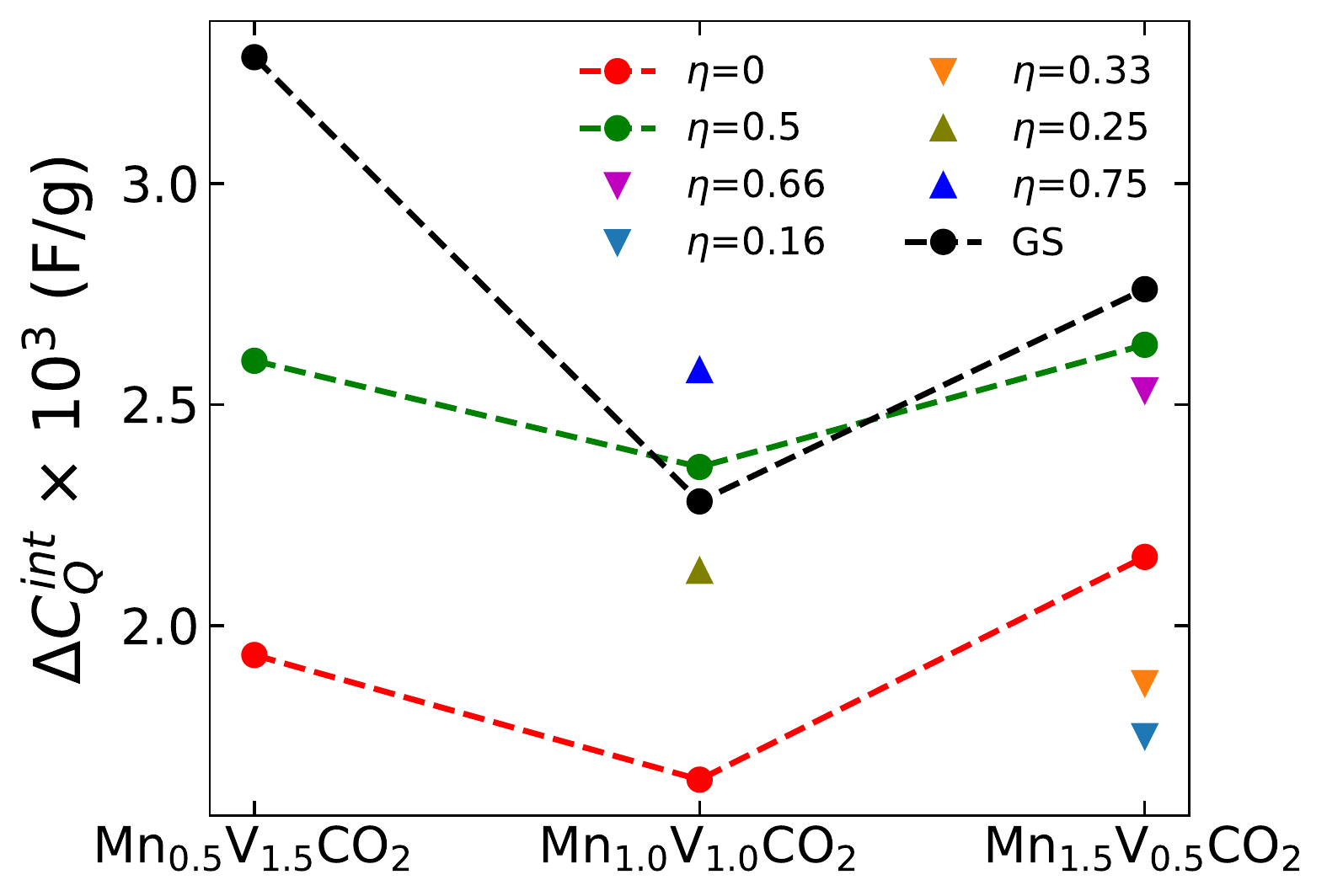}
    \caption{$\Delta C_{Q}^{int}$ for ground and all other $\eta$ states of V$_{2-x}$Mn$_x$CO$_2$; x=0.5, 1.0 and 1.5 }
    \label{Fig:9}
\end{figure}
For all compositions, there is no specific magnetic state that can be singled out to be the provider of the highest value of $C_{Q}^{int}$. In fact on the positive side of the voltage window, several partially ordered magnetic states along with the ground states have nearly the same maximum value of $C_{Q}^{int}$. Due to the lack of any clear trend over the entire voltage window, a more appropriate quantity to assess the potential best choice among these compounds is $\Delta C_Q^{int}$, the difference between the maximum and minimum values of $C_Q^{int}$. The one with the least $\Delta C_Q^{int}$ is expected to be a better choice as one can then work with a larger voltage window. In  Figure \ref{Fig:9}, we show $\Delta C_Q^{int}$ of V$_{2-x}$Mn$_{x}$CO$_2$ compound for the magnetically ordered ground state and various spin-disordered states associated with each value of $x$. The results show that the variations in the quantum capacitance over the entire voltage window are largest for the ground states of Mn-deficient and Mn-rich compounds, while the least variations are found for fully or near fully spin-disordered states. This implies that using these MXenes as electrodes can be advantageous if operated at higher temperatures.

\begin{table*}
\setlength{\tabcolsep}{8pt} 
\renewcommand{\arraystretch}{1.0} 
    \caption{Total charge transfer($\Delta Q_{tot}$), Change in work function($\Delta WF$), Redox Capacitance(C$_{redox}$), EDL Capacitance(C$_{EDL}$) and Electrical Capacitance(C$_{E}$) of different magnetic states for V$_{2-x}$Mn$_{x}$CO$_{2}$}
    \begin{tabular}{|c@{\hspace{0.5cm}}| c@{\hspace{0.5cm}} | c@{\hspace{0.5cm}} | c@{\hspace{0.5cm}} |c@{\hspace{0.5cm}} |c@{\hspace{0.5cm}}|} 
    \hline
    \vspace{-0.33cm}
    \\ System & states & $\Delta Q_{tot}$ ($e$) & $\Delta WF$ ($eV$) & C$_{redox}$ (F/g) & C$_{E}$ (F/g) \\
    \hline
     V$_{1.5}$Mn$_{0.5}$CO$_{2}$ & GS(AFM-c) & 14.19 & 5.25 & 110.25  & 166.03 \\
                                 & $\eta$ = 0.50 & 14.05 & 5.22 & 109.81  & 165.59 \\
                                 & $\eta$ = 0.00 & 14.09 & 5.29 & 108.76 &  164.54\\
    \hline    
    V$_{1.0}$Mn$_{1.0}$CO$_{2}$ & GS(FM) & 14.25 & 5.49 & 104.29  & 160.99 \\
                                & $\eta$ = 0.75 & 14.26 & 5.41 & 105.91  & 162.61 \\
                                & $\eta$ = 0.50 & 14.27 & 5.44 & 105.40  & 162.10 \\
                                & $\eta$ = 0.25 & 14.28 & 5.52 & 103.94  & 160.64 \\
                                & $\eta$ = 0.00 & 14.29 & 5.47 & 104.96  &  161.66\\
    \hline   
    V$_{0.5}$Mn$_{1.5}$CO$_{2}$ & GS(FeM) & 16.00 & 5.35 & 118.58 & 172.76 \\
                                & $\eta$ = 0.66 & 14.13 & 5.70 & 98.29  & 152.47 \\
                                & $\eta$ = 0.50 & 14.14 & 5.81 & 96.67  & 150.85 \\
                                & $\eta$ = 0.33 & 14.12 & 5.85 & 95.70  & 149.88 \\
                                & $\eta$ = 0.16 & 14.14 & 5.88 & 95.35  & 149.53 \\
                                & $\eta$ = 0.00 & 14.22 & 6.11 & 92.28  & 146.46\\
    \hline                            
    \end{tabular}
    \label{Tab:2}
\end{table*}    

\begin{figure}[ht!]
    \centering
    \includegraphics[width=0.7\linewidth]{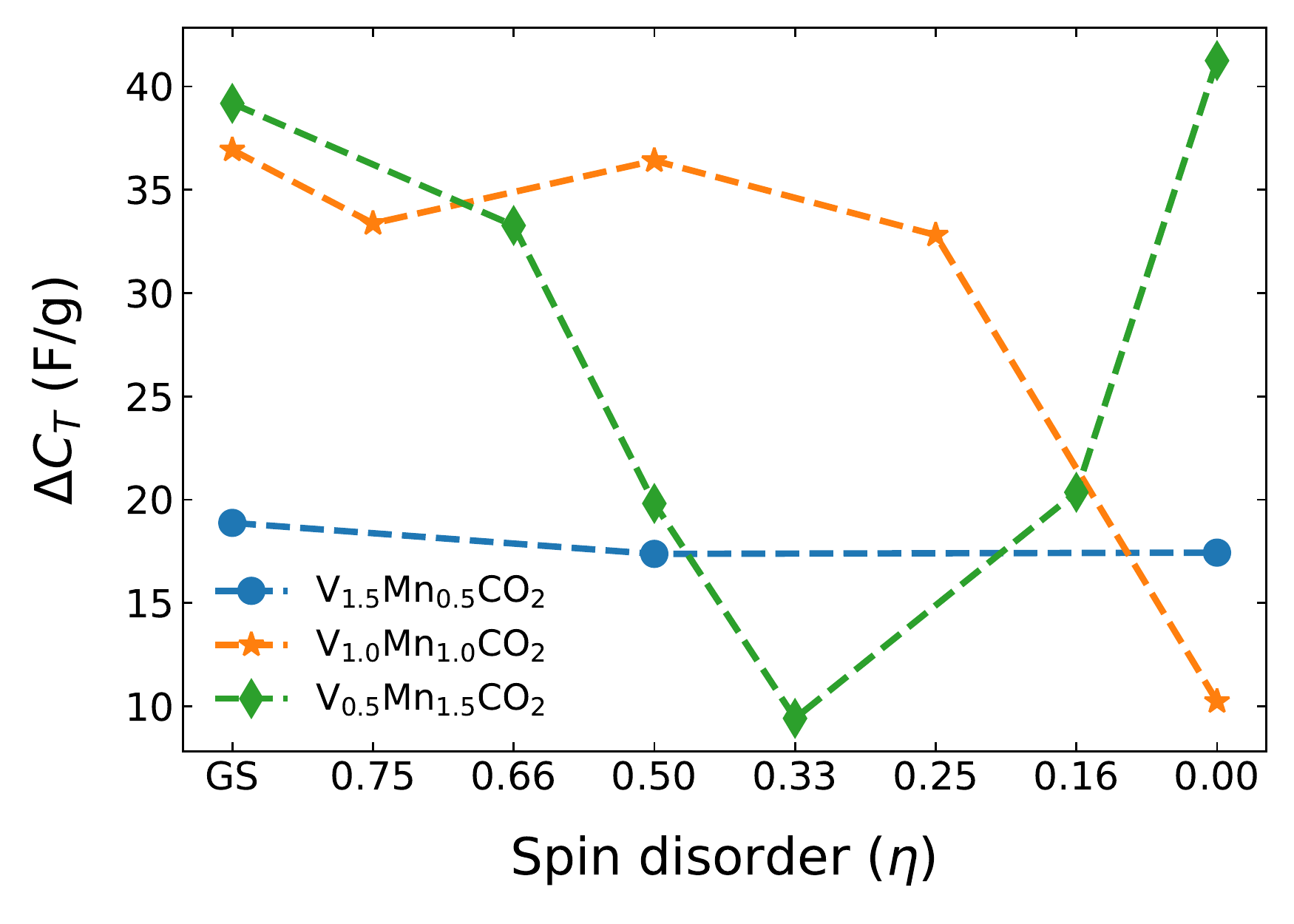}
    \caption{$\Delta$C$_{T}$ for the ground state (GS) and different spin disorder ($\eta$) of V$_{2-x}$Mn$_{x}$CO$_{2}$(x=0.5,1.0,1.5) }
    \label{Fig:11}
\end{figure}

The redox (C$_{redox}$) and electrochemical double layer (C$_{EDL}$) capacitances are calculated using equations (\ref{EQN: 6})-(\ref{EQN: 9}). In the formalism used, the variations in the EDL capacitance are only due to the changes in the area of the electrode, which is estimated by the lattice parameter of the compound, as the other parameters ($\epsilon_0$, $\epsilon_r$ =6 and $d$ = 2.8 $\AA$) are constant for a given electrolyte. The calculated EDL capacitance of V$_{1.5}$Mn$_{0.5}$CO$_2$, V$_{1.0}$Mn$_{1.0}$CO$_2$, V$_{0.5}$Mn$_{1.5}$CO$_2$ is 55.78F/g, 56.70 F/g and 54.18 F/g, respectively. The surface redox contribution to the electrical capacitance, on the other hand, is expected to be significant with changes in chemical and magnetic disorder as different chemical and magnetic constitutions at the surfaces along with variations in bond distances should affect the surface charge transfer. In Table \ref{Tab:2} C$_{redox}$ and C$_E$)(Equation \ref{EQN: 9}) values along with the total charge transfer ($\Delta Q_{tot}$) through surfaces and changes in the work functions ($\Delta {WF}$) are shown for different magnetic states and chemical compositions. The calculated values of these quantities demonstrate the following: (a) There is no significant change in C$_{redox}$ and thus in C$_{E}$ values of V$_{1.5}$Mn$_{0.5}$CO$_2$ and V$_{1.0}$Mn$_{1.0}$CO$_2$ as one changes the magnetic state from the ordered ground state to the complete spin disordered state. Thus, neither the chemical composition nor the spin disorder has any substantial effect on the electrical capacitances of these two compounds. The insensitivity in these quantities with regard to disorders is due to a near constant charge transfer and a maximum variation of 4$\%$ in $\Delta WF$, (b) the case of Mn-rich composition V$_{0.5}$Mn$_{1.5}$CO$_2$ shows an anomaly. Here both $\Delta Q_{tot}$ and $\Delta WF$ for the ground state are significantly different from the values obtained for various magnetically disordered states of this particular compound. A larger charge transfer and a relatively smaller $\Delta WF$ in the ground state produces a $C_{E}$ value of 172 F/g, the highest among all cases considered. Moreover, $C_{redox}$ for various $\eta$ states are lower among all systems across compositions and magnetically disordered states. This is due to the fact that upon introduction of spin disorder in this compound $\Delta Q_{tot}$ decreases by about 2e; e the charge of the electron, and stays near constant with changes in the degree of spin disorder. Moreover, the $\Delta Q_{tot}$ values for the spin-disordered states are almost the same as those found in the other two compounds. On the other hand, $\Delta WF$ values are relatively high in comparison with those in the other two compounds along with a $7\%$ variation among the magnetically disordered states. This variation is only $1\%$ for the other two compounds. The root of this anomalous behavior lies in the magnetic constitution of the two MXene surfaces and will be discussed in the next subsection.    
\begin{figure*}
    \centering
    \begin{subfigure}[b]{0.30\linewidth}
    \includegraphics[width=\linewidth]{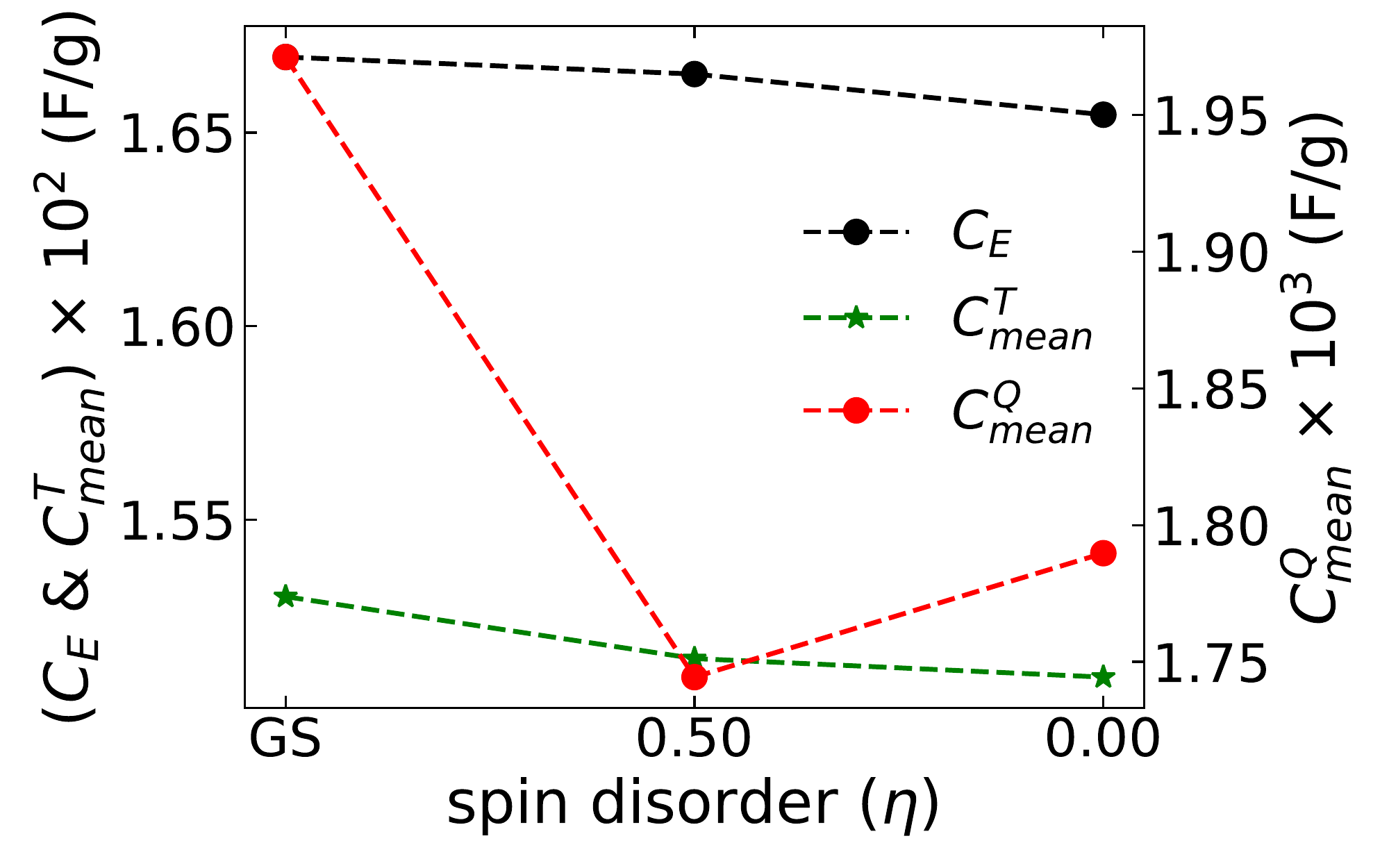}
    \caption{V$_{1.5}$Mn$_{0.5}$CO$_{2}$}
    \end{subfigure}
    \hspace{-0.00cm}
    \begin{subfigure}[b]{0.30\linewidth}
    \includegraphics[width=\linewidth]{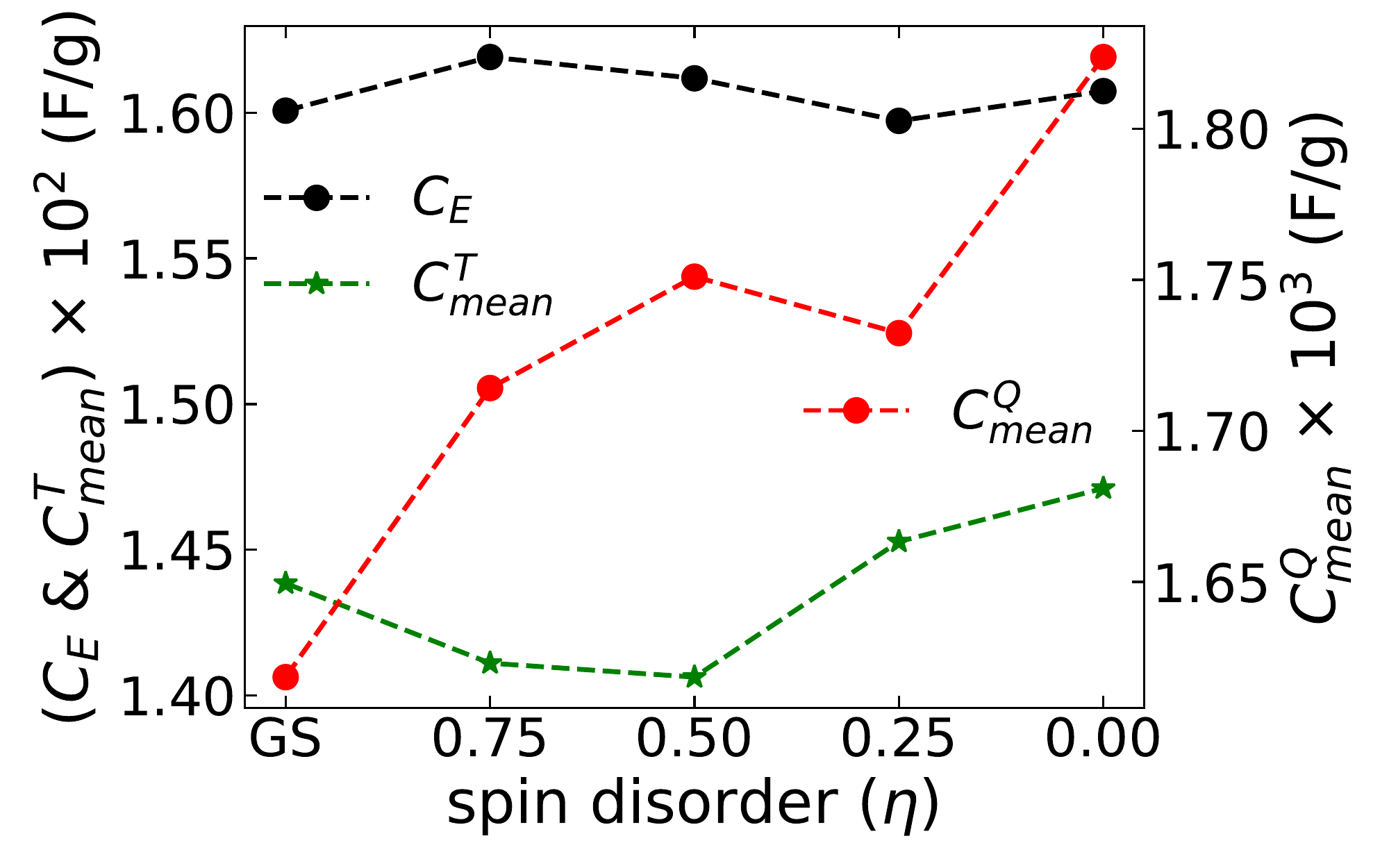}  
    \caption{V$_{1.0}$Mn$_{1.0}$CO$_{2}$}
    \end{subfigure}
    \hspace{-0.00cm}
    \begin{subfigure}[b]{0.3\linewidth}
    \includegraphics[width=\linewidth]{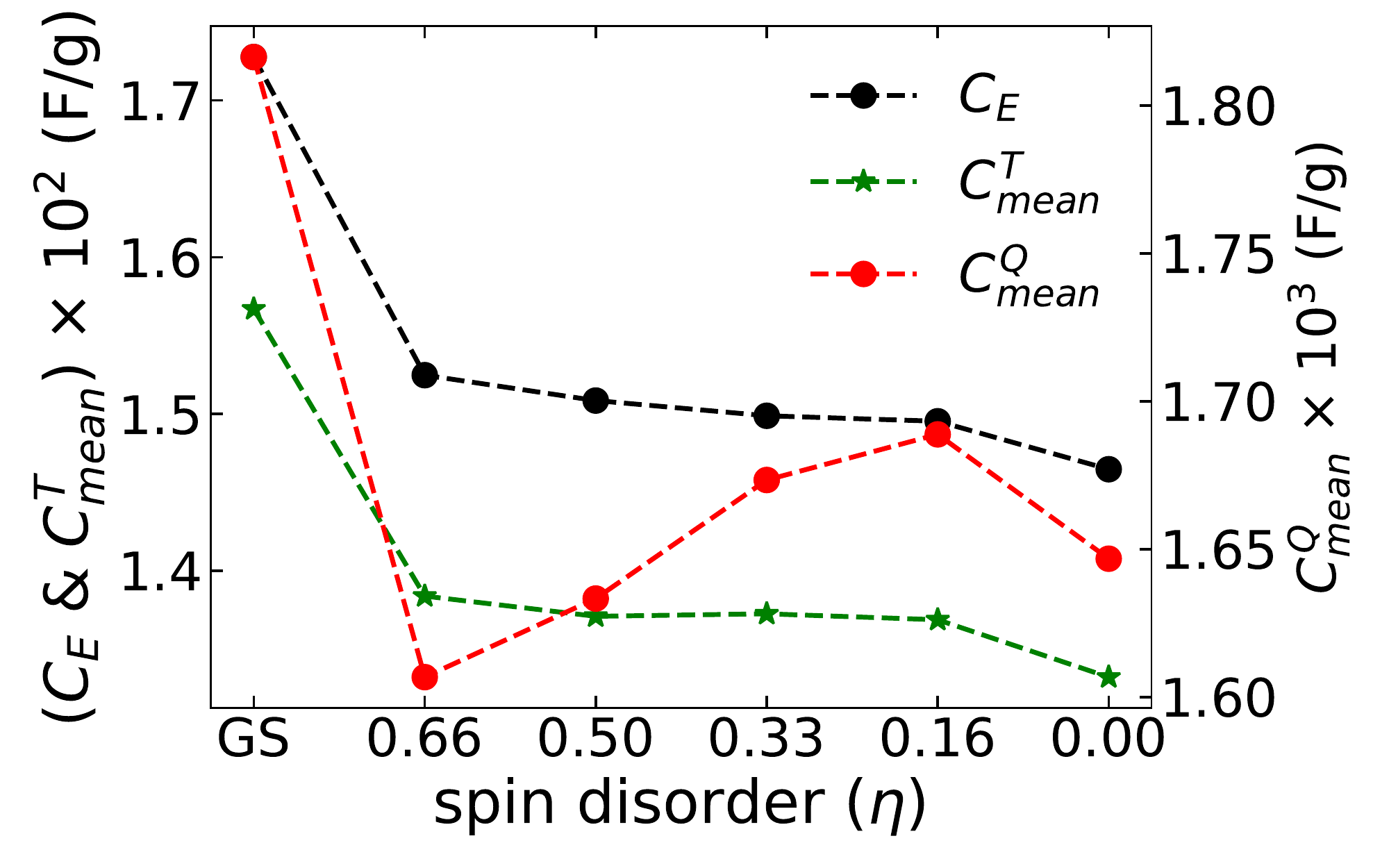}
    \caption{V$_{0.5}$Mn$_{1.5}$CO$_{2}$}
    \end{subfigure}
    \caption{C$_{E}$, C$^{T}_{mean}$ and C$^{Q}_{mean}$ for V$_{2-x}$Mn$_{x}$CO$_{2}$, where $x$ = 0.5, 1.0 and 1.5.}
    \label{Fig:10}
\end{figure*}
The total capacitance is calculated using equation \ref{EQN: 2}. However, since $C_{Q}$ is a function of the applied voltage while $C_{E}$ is not, it is more appropriate to consider the average values of $C_{Q}$ and hence $C_{T}$ for each case. The averages of quantum($C_{mean}^{Q}$) and total capacitance ($C_{mean}^{T}$) are calculated by taking the averages over the entire potential window considered here. The average values of the capacitances imply the most likely outcome of the experiments done within the voltage window. In Figure \ref{Fig:10}, we present $C_E$, $C_{mean}^{Q}$ and $C_{mean}^{T}$ as a function of the degree of spin disorder $\eta$ for each one of the three compounds. We find that in all three compounds, the variations in the $C_{mean}^{T}$ follow the trends in the $C_{E}$ closely. This is understandable from the fact that $C_{mean}^{Q}$ is one order of magnitude smaller than $C_{E}$, and thus its role of it primarily is to reduce the total capacitance from its electrical capacitance value. The results demonstrate that while the $C_{mean}^{T}$ in V$_{1.5}$Mn$_{0.5}$CO$_{2}$ varies insignificantly with the spin disorder, the most significant variation is found in V$_{0.5}$Mn$_{1.5}$CO$_{2}$. The variations of $C_{mean}^{T}$ in V$_{1.0}$Mn$_{1.0}$CO$_{2}$ is non-monotonic. The highest $C_{mean}^{T}$ is obtained in the ground state of the Mn-rich compound.
In order to decide on the potential best choice for an electrode that has minimum fluctuation across the voltage window, we compute $\Delta C_T$, the difference between the maximum and minimum values of $C_{T}$ for each chemical and magnetic composition. The results are presented in Figure \ref{Fig:11}. The V-rich system has a low $\Delta C_T$ that is almost insensitive to the magnetic state. This means that it can be operated over a large range of temperatures. For the other two compounds, $\Delta C_T$ values lower than this are obtained. However, in order to exploit this trait, they have to be operated at specific spin-disordered states, that is at specific temperatures which can be considerably high. This poses a disadvantage in using V$_{1.0}$Mn$_{1.0}$CO$_{2}$ and V$_{0.5}$Mn$_{1.5}$CO$_{2}$ as electrodes in supercapacitors.   

\subsection{Anomalous electrochemical behaviour of V$_{0.5}$Mn$_{1.5}$CO$_{2}$}

\begin{figure}[ht!]
    \centering
    \includegraphics[width=1.0\linewidth]{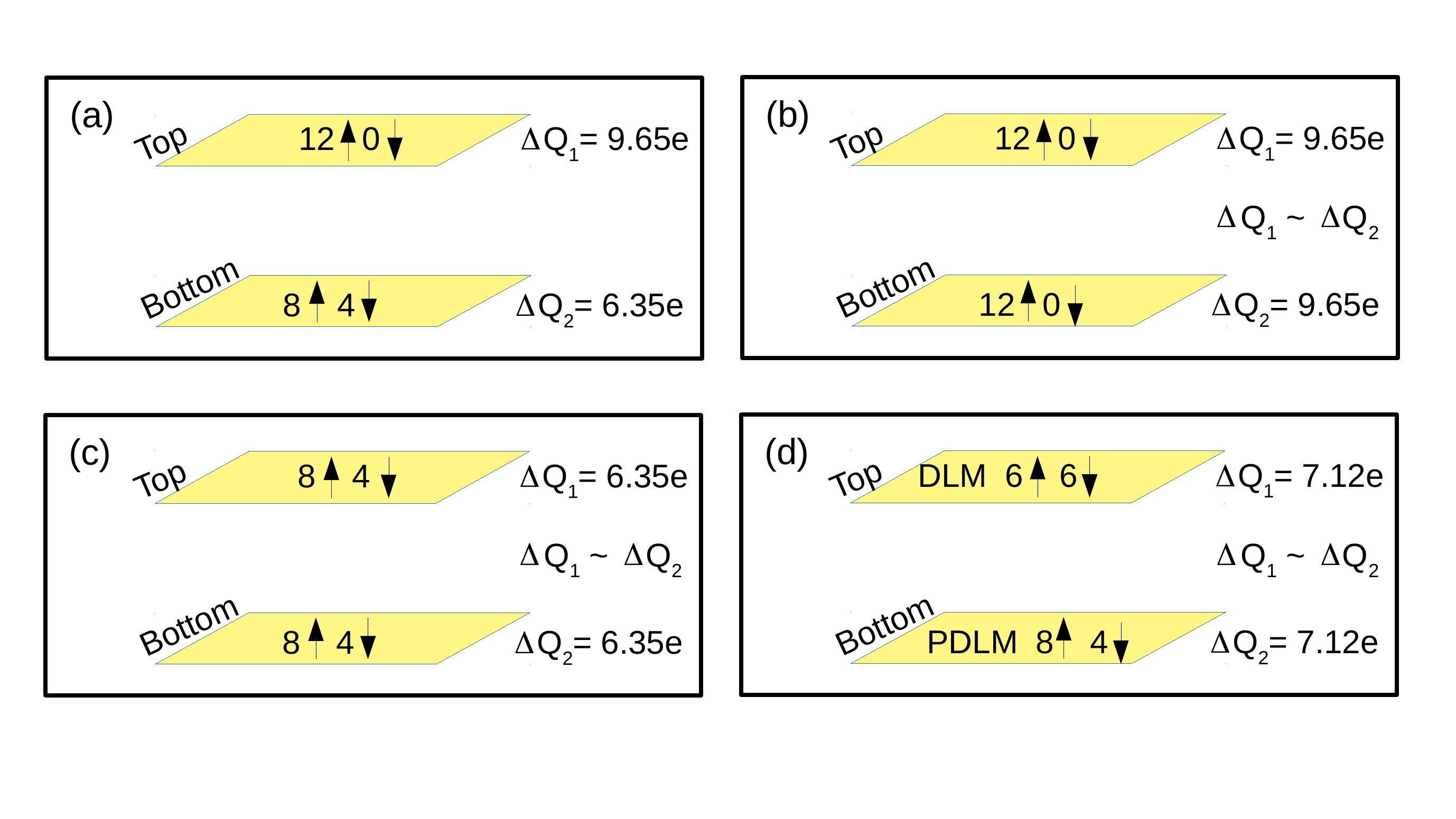}
    \caption{Schematic diagram showing various different magnetic structures on the top and bottom surface of V$_{0.5}$Mn$_{1.5}$CO$_2$ and corresponding charge transfer from them. (a) The ground state, (b) a situation where spins of all Mn atoms (top and bottom surface ) are aligned along the c-axis, a replica of the top surface in the ground state, (c) a situation where spins of 8 Mn(4 Mn) atoms on both surfaces are aligned(anti-aligned) along the c-axis, a  replica of the bottom surface in the magnetic ground state. (d) a situation where the top surface has a spin arrangement consistent with $\eta$=0.00 (complete spin disordered state) and the bottom surface spin arrangement is according to the partial magnetic disordered state denoted by $\eta$=0.33.    }
    \label{Fig:12}
\end{figure}

In order to understand the origin of the anomalous electrochemical behavior of the Mn-rich system,$i.e$, V$_{0.5}$Mn$_{1.5}$CO$_{2}$, we perform a detailed analysis of the charge transfer through the two surfaces. The ground state surface magnetic structure of this compound is unique in the sense that, unlike the other two compounds, in this case, the spin orientations of Mn atoms on the two surfaces are different. In Table \ref{tab:1}, Appendix B, we show the total charge transfers from both surfaces for all compounds in different magnetic states. We find that except for the ground state of the Mn-rich compound, the charge transfer from either surface, irrespective of the chemical composition and magnetic state, denoted by $\eta$ value, is the same. For the system in question, there is about 2.5e(1.5e) more(less) charge transfer from the top(bottom) surface in comparison to that for all other systems. To get more insight into this anomalous behavior, we have computed the Bader charges \cite{bader} on Mn, V, and O atoms of both surfaces of V$_{0.5}$Mn$_{1.5}$CO$_{2}$ before and after hydrogenation. The results are presented in Tables \ref{tab:2} and \ref{tab:3} of Appendix B. The results are shown for four non-equivalent atom clusters only.
From the computed Bader charges, we find that before hydrogenation, all V atoms had an identical charge of 11.07e irrespective of which surface they occupy. After hydrogenation, there is a 0.09e charge more on the V atoms on the top surface. The Mn and C atoms are almost inert in the redox charge transfer through hydrogenation of the surfaces as the changes in the Mn (C) charges after hydrogenation are between 0.04e-0.07e (0.09e-0.12e) only. As expected the surface charge transfer happens through O atoms and the largest changes in charge content happen for them. However, the results suggest that the amount of change depends on the location of the O atoms. The charge content on the O atoms on the top surface, as a result of hydrogenation, changes between 0.4e-0.52e while for those on the bottom surface, it is between 0.2e-0.34e. The reason behind more charge transfer through the V atoms could be the rumpling structure of the surfaces in this compound. The rumpling pulls V surfaces closer to the O and thus the H atoms. This is clear from the V-O bond distances that vary between 1.72-1.83 $\AA$ whereas the Mn-O bond distances have a wider dispersion (1.92-2.11 $\AA$).
The substantially different charge transfer from two surfaces can be correlated with the magnetic environment around an atom on each surface in the ground state of V$_{0.5}$Mn$_{1.5}$CO$_{2}$. Among the O atoms in the bottom surface, O$_{21}$-O$_{24}$ (brown block in Figure \ref{fig:15}(b), Appendix B) transfer only 0.17e upon hydrogenation. Each one of these four is networked with an Mn atom which has its spin aligned along $-c$-direction. The same is true for O$_{25}$-O$_{28}$ (magenta block in Figure \ref{fig:15}(b), supplementary information) where only 0.19e electrons are transferred. On the other hand, O$_{17}$-O$_{20}$ (green block in Figure \ref{fig:15}(b), Appendix B) are networked with 3 Mn out of which 2 have spins aligned along $c$-direction, the associated charge transfer is 0.33e. A charge transfer of 0.3e happens in case of O$_{29}$-O$_{32}$ which are in a network with 2 Mn, both having spins aligned along $c$-direction. Therefore, the presence of at least two Mn having their spins aligned in the same direction enables more charge transfer through the O atoms. All Mn spins on the top surface are aligned in the same direction and the charge transfers associated with O atoms are more. The differences in the amounts of charge transfer for different groups of atoms can be explained through the variations in the transition metal neighborhood of the O atoms. A charge transfer of 0.4e occurs for atoms O$_{1}$-O$_{4}$ (orange block in Figure \ref{fig:15}(a), Appendix B). These O have three Mn atoms in their neighborhood. The numbers are 0.4-0.42e for O$_{5}$-O$_{12}$. Here the O is in a network of two Mn and one V atoms (blue and black blocks in Figure \ref{fig:15}(a), Appendix B) with Mn-O (V-O) bond distances of about 2.0 $\AA$(1.8 $\AA$). The largest charge transfer of 0.52e is associated with the remaining four oxygen (red block in Figure \ref{fig:15}(a), Appendix B) which have the same neighborhood of transition metals as O$_{5}$-O$_{12}$ but the V-O distances are shorter (1.72 $\AA$). 
The correlation of charge transfer with the magnetic structure on the surface is, therefore, clear. Since there is identical charge transfer from either surface in partial or fully disordered magnetic states for this compound, it appears that the non-uniformity in spin distribution on the chemically identical surfaces must be responsible for non-uniformity in the charge transfer. We have further examined this proposition by calculating the charge transfer from the surfaces by various combinations of magnetic configurations on the surfaces of V$_{0.5}$Mn$_{1.5}$CO$_{2}$. In Figure \ref{Fig:12} the situations considered are shown schematically along with the charge transfer $\Delta Q_{1}(\Delta Q_{2})$ from the top(bottom) surfaces. The ground state is shown in Figure \ref{Fig:12}(a) while two different situations with identical ordered configurations on both surfaces are shown in Figure \ref{Fig:12} (b)-(c). In Figure \ref{Fig:12}(b), both surfaces are ferromagnetic, the replica of the top surface in the ground state. In Figure \ref{Fig:12}(c), both surfaces are replicas of the bottom surface in the ground state. In Figure \ref{Fig:12}(d), a completely different situation with a top surface having a complete spin disordered configuration ($\eta=0$) and a bottom surface with a partial spin disordered configuration ($\eta=0.33$) is depicted. It is to be noted that the number of spins aligned(anti-aligned) along(against) $c$-axis in $\eta=0.33$ state is the same as that in the bottom surface in the ground state. When both surfaces are ferromagnetic, the charge transfer from both surfaces is identical and exactly the same as that from the top surface when the system is in the ground state (Figure \ref{Fig:12}). In situation (c), we find identical charge transfer from both surfaces with a magnitude identical to that corresponding to the bottom surface in the ground state which has an identical spin configuration. From these results, two things appear to emerge. First, when the surfaces are identical in terms of the magnetic environment around an atom, the charge transfers are identical, and second, a ferromagnetic configuration promotes higher charge transfer while a ferrimagnetic ordered configuration promotes lower charge transfer in comparison with those in magnetically disordered states. 
The results obtained from the situation depicted in Figure \ref{Fig:12}(d), however, demonstrate that identical magnetic configuration on the surfaces is not a necessary condition for identical charge transfer from both surfaces. The charge transfer numbers in this configuration nevertheless confirm that when there is a mixture of aligned and anti-aligned spins (along $c$-direction), a random distribution leads to higher charge transfer. Since the chemical environment and structure around an atom are kept identical in all four situations discussed, the differences in charge transfer must be a consequence of different magnetic environments around an atom. In the ordered ferrimagnetic configuration considered here (Figure \ref{Fig:12}(c)), the environment around each atom follows a particular order but in the partially ordered situation denoted by $\eta=0.33$, the order vanishes. That the ferromagnetic order promotes the highest charge transfer, is, however, seen only with this chemical composition. In the other two compounds, despite the intra-surface order being ferromagnetic, we do not find charge transfer of this large amount. Therefore, the higher concentration of Mn should be the reason. The results, thus, imply the importance of both chemical and magnetic composition.     
\section{Conclusion}
Using DFT-based first-principles electronic structure calculations, we have explored in detail the effects of chemical and magnetic disorder on the capacitance of solid solution MXene V$_{2-x}$Mn$_{x}$CO$_{2}$ electrode in an electrode-acidic electrolyte supercapacitor setup. Although the material is yet to be exfoliated from its precursor MAX phase, we have considered this as a test case to address the yet unexplored issues of disorder effects on its charge storage capacity. We have systematically computed various contributions to the total capacitance with changes in the chemical composition and magnetic structure. We find that (a) the ground state magnetic structure is dependent upon the chemical composition (b) the total capacitance does not vary significantly with changes in chemical and magnetic order (c) the total capacitance of this material is comparable to the capacitance of Ti$_{3}$C$_{2}$O$_{2}$, the most extensively studied MXene for supercapacitor application and (d) the minimum and constant variation in the total capacitance with the variation of spin disorder is obtained for V$_{1.5}$Mn$_{0.5}$CO$_{2}$, a trait that is most useful for operational purpose. In the course of our investigation, we have also found anomalous behavior in surface charge transfer. The asymmetric magnetic configurations on two surfaces lead to significantly different charge transfers from different surfaces affecting the redox capacitance. We find that the inhomogeneity in the magnetic structure across surfaces affects the charge transfer mutually. The most important finding is that in this composition only, ferromagnetic configuration promotes a significantly high charge transfer and thus a substantially large capacitance can be achieved by designing the system with both surfaces having ferromagnetic order. This is a new phenomenon hitherto not reported in the literature. This can have a profound effect on tuning the capacitance of the electrode by manipulating the magnetic structure of the MXene surfaces. However, a microscopic understanding of the phenomenon as well as the reason behind different magnetic ground states for different chemical compositions of (V-Mn)CO$_{2}$ MXene are beyond the scope of this work and will be taken up in the future.
\appendix
\renewcommand{\thefigure}{A\arabic{figure}}
\setcounter{figure}{0}
\renewcommand{\thetable}{A\arabic{table}}
\setcounter{table}{0}
\begin{appendix}
\section{Chemical arrangement and electronic structure}
\begin{figure}[h!]
    \centering
    \begin{subfigure}[b]{0.25\linewidth}
    \includegraphics[width=\linewidth]{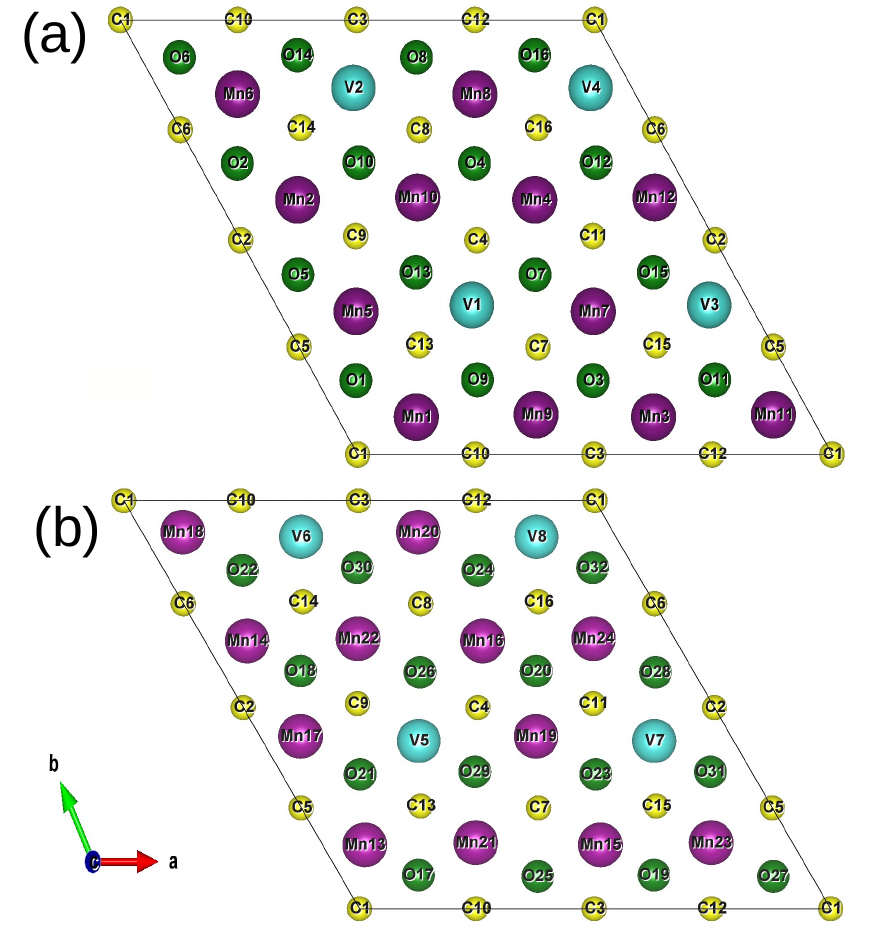}
    \end{subfigure}
    \hspace{-0.00cm}
    \begin{subfigure}[b]{0.26\linewidth}
    \includegraphics[width=\linewidth]{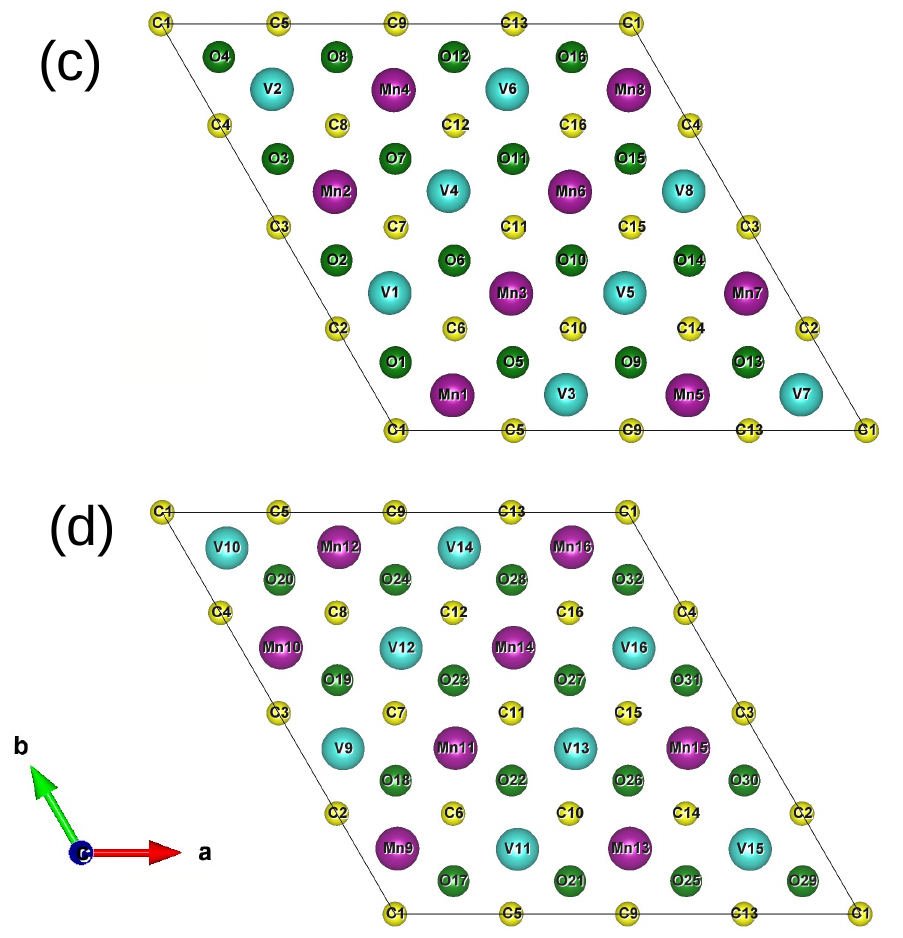}  
    \end{subfigure}
    \hspace{-0.00cm}
    \begin{subfigure}[b]{0.28\linewidth}
    \includegraphics[width=\linewidth]{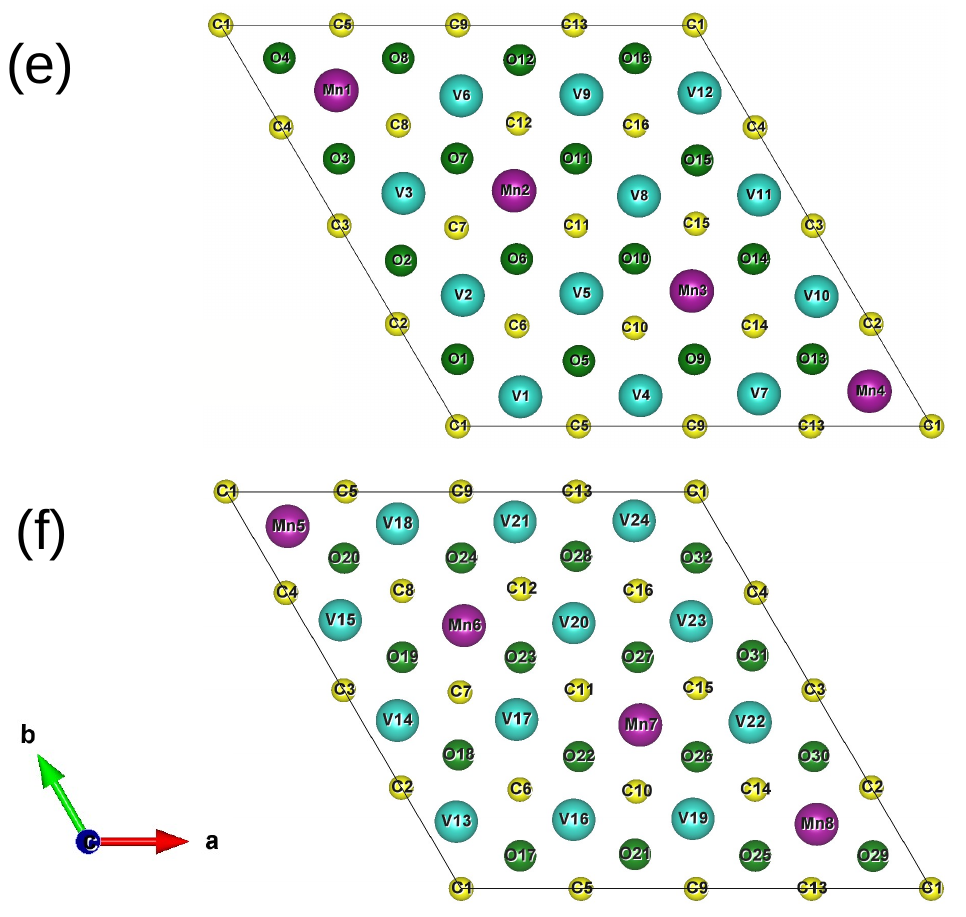}
    \end{subfigure}
    \caption{(a),(c),(e) are the top layer and (b),(d),(f) are the bottom layer chemical arrangement of V$_{0.5}$Mn$_{1.5}$CO$_2$, V$_{1.0}$Mn$_{1.0}$CO$_2$ and V$_{1.5}$Mn$_{0.5}$CO$_2$.}
    \label{fig:1}
\end{figure}

\begin{figure}[ht!]
    \centering
    \includegraphics[width=0.7\linewidth]{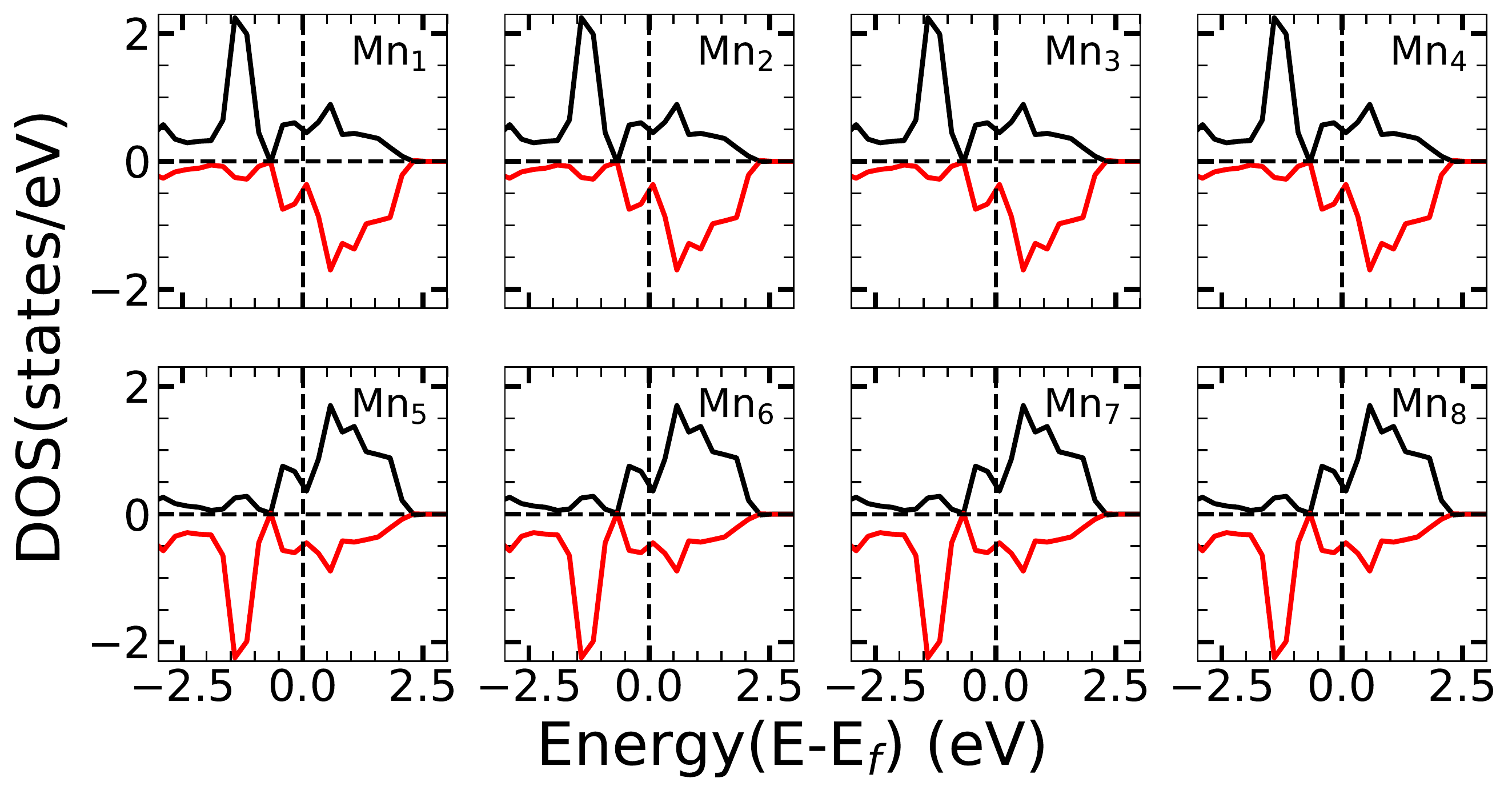}
    \caption{Projected Densities of States each Mn atoms in Ground State of V$_{1.5}$Mn$_{0.5}$CO$_2$}
    \label{fig:2}
\end{figure}

\begin{figure}[ht!]
    \centering
    \includegraphics[width=0.7\linewidth]{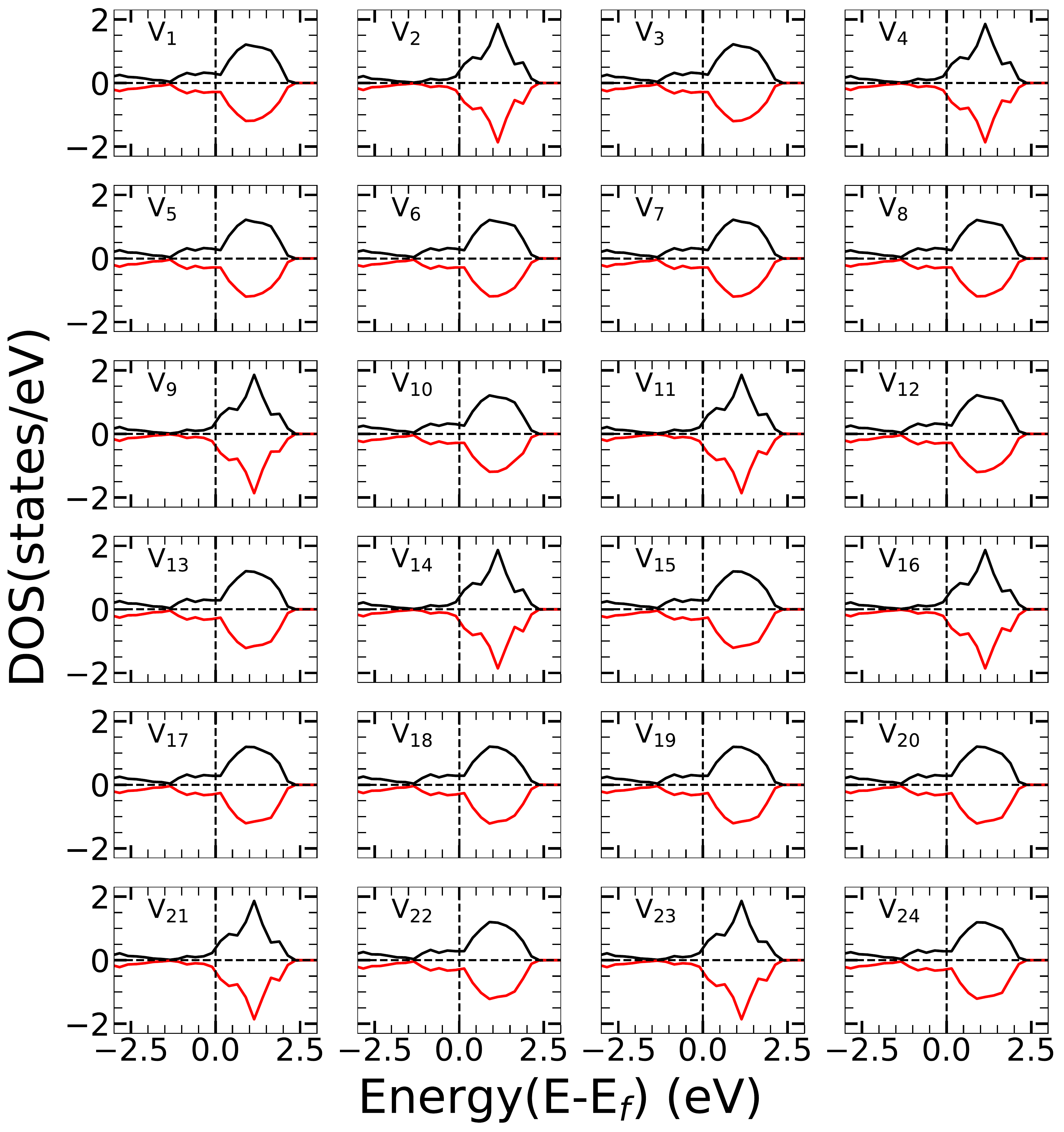}
    \caption{Projected Densities of States each V atoms in Ground State of V$_{1.5}$Mn$_{0.5}$CO$_2$}
    \label{fig:3}
\end{figure}

\begin{figure}[H]
    \centering
    \includegraphics[width=0.7\linewidth]{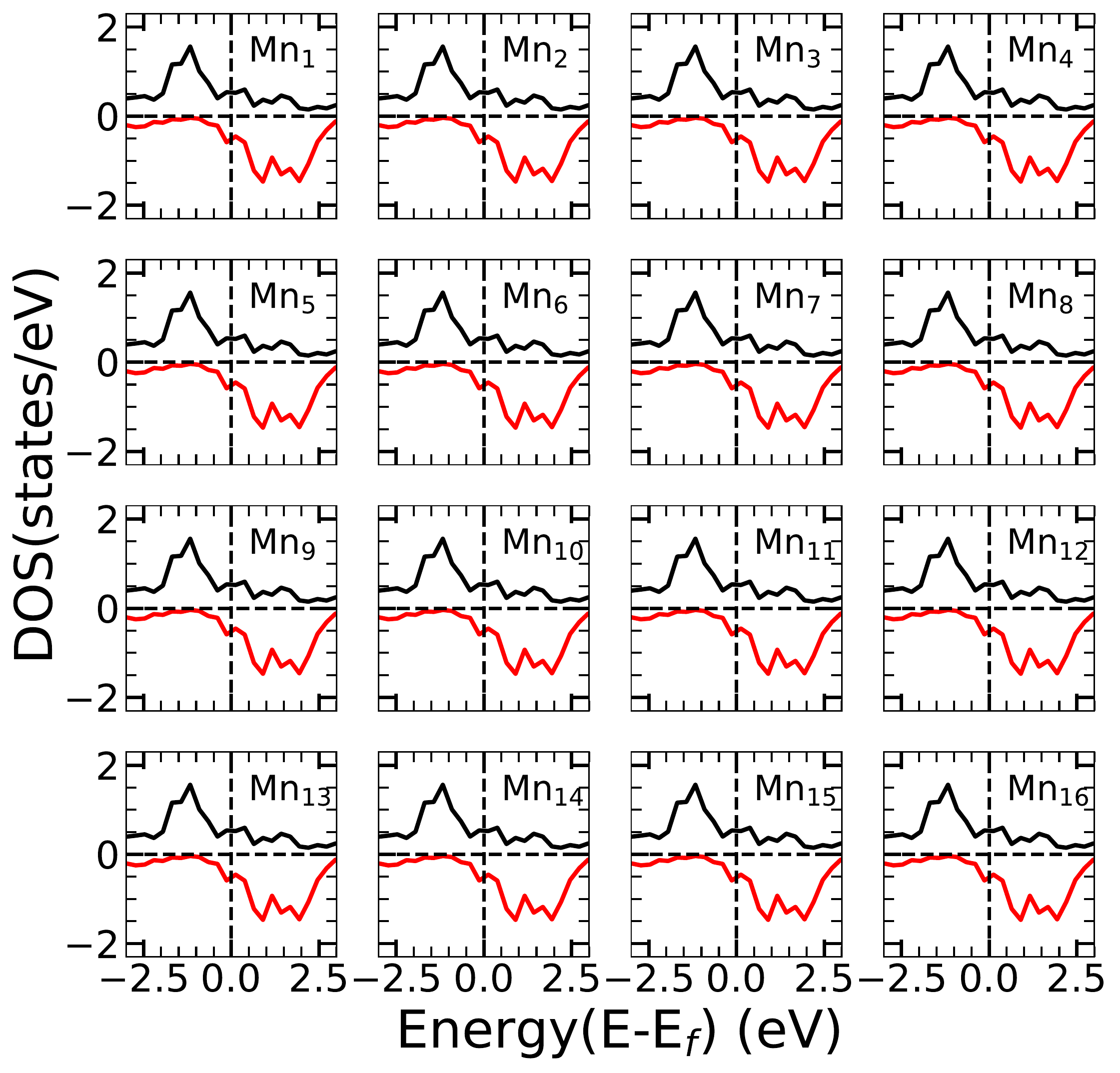}
    \caption{Projected Densities of States each Mn atoms in Ground State of V$_{1.0}$Mn$_{1.0}$CO$_2$}
    \label{fig:4}
\end{figure}

\begin{figure}[H]
    \centering
    \includegraphics[width=0.7\linewidth]{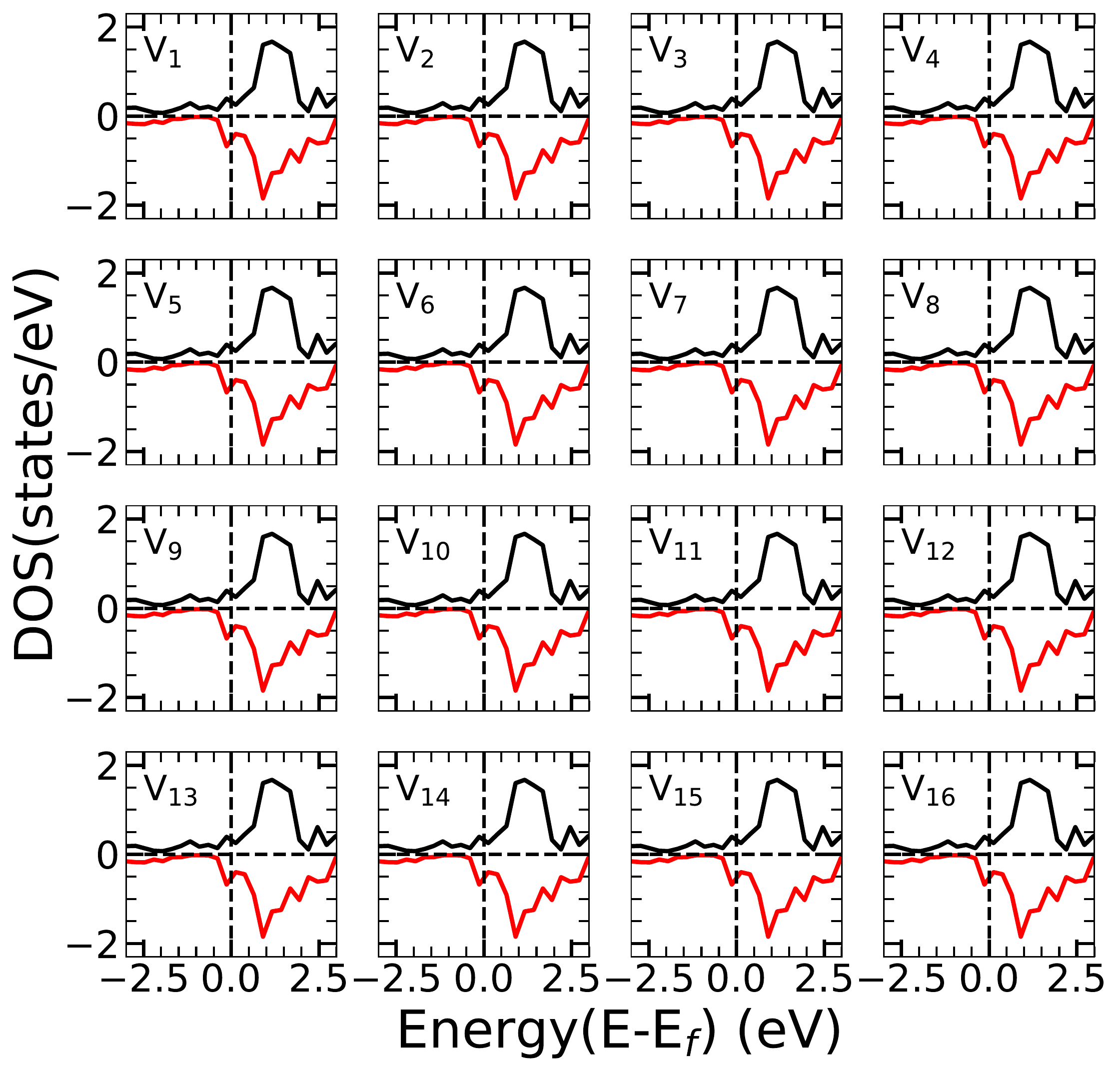}
    \caption{Projected Densities of States each V atoms in Ground State of V$_{1.0}$Mn$_{1.0}$CO$_2$}
    \label{fig:5}
\end{figure}

\begin{figure}[H]
    \centering
    \includegraphics[width=0.7\linewidth]{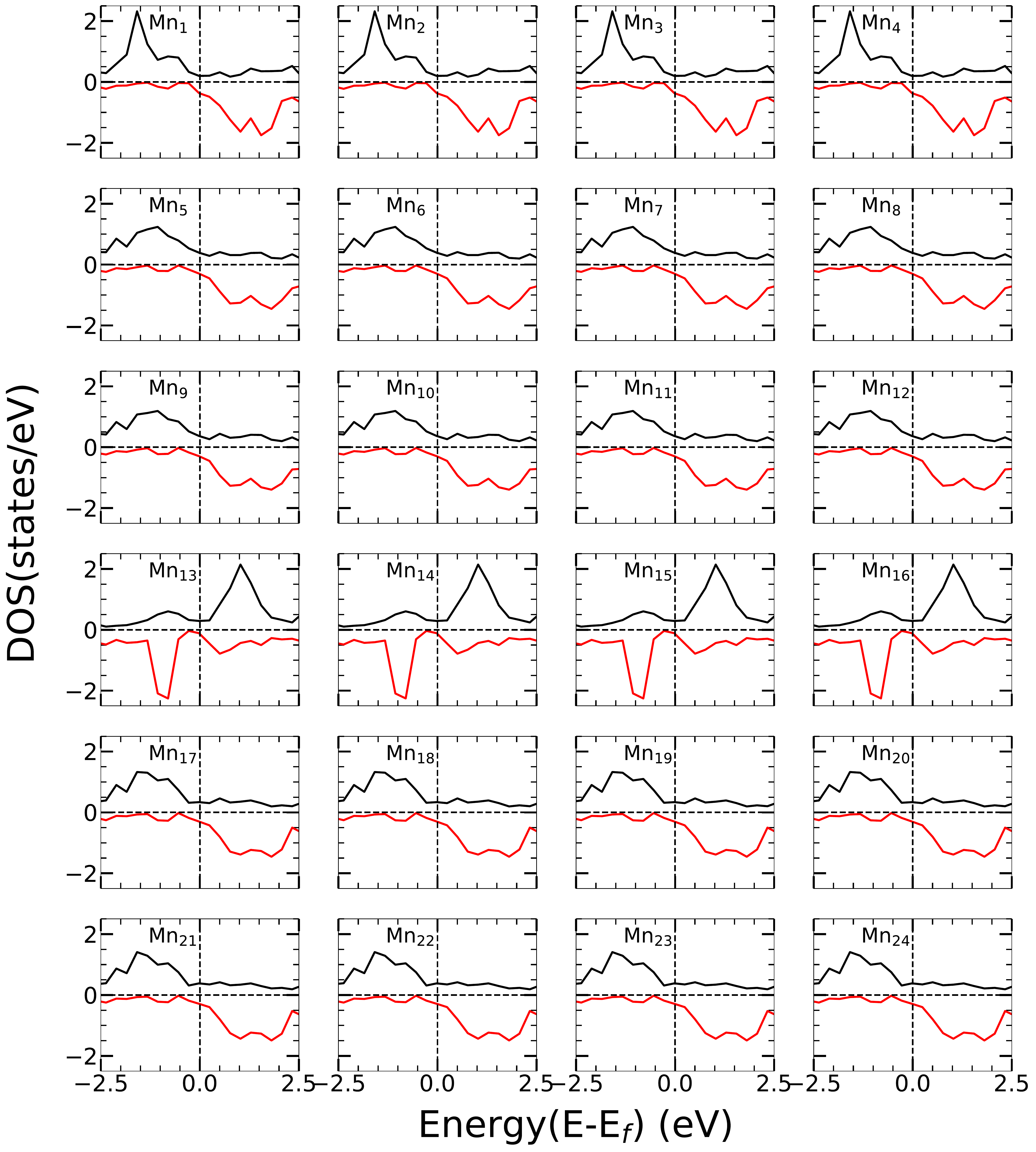}
    \caption{Projected Densities of States each Mn atoms in the Ground State of V$_{0.5}$Mn$_{1.5}$CO$_2$}
    \label{fig:6}
\end{figure}

\begin{figure}[H]
    \centering
    \includegraphics[width=0.7\linewidth]{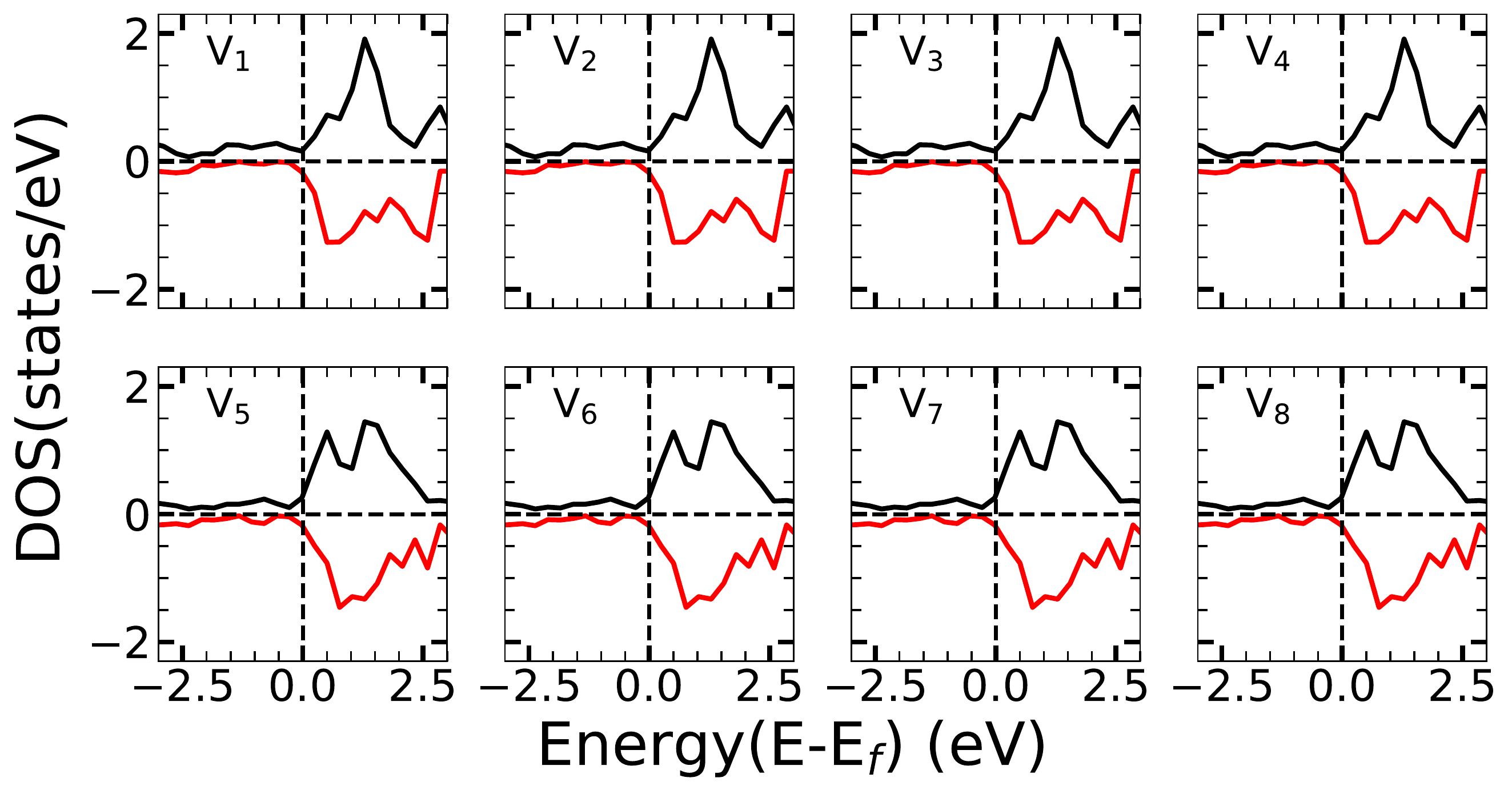}
    \caption{Projected Densities of States each V atoms in the Ground State of V$_{0.5}$Mn$_{1.5}$CO$_2$}
    \label{fig:7}
\end{figure}

\begin{figure}[H]
    \centering
    \includegraphics[width=0.7\linewidth]{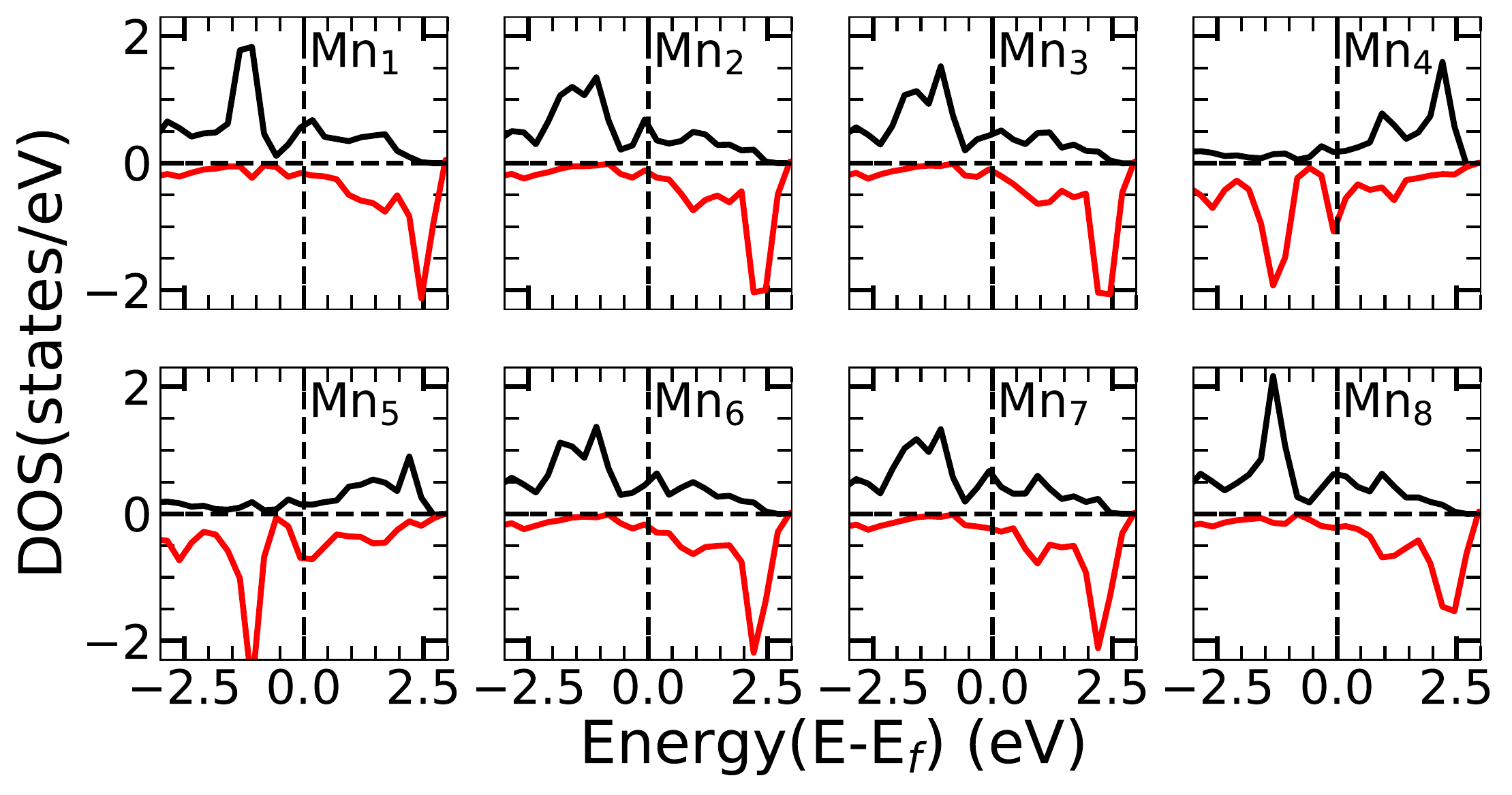}
    \caption{Projected Densities of States each Mn atoms in $\eta$=0.50 of V$_{1.5}$Mn$_{0.5}$CO$_2$}
    \label{fig:8}
\end{figure}

\begin{figure}[H]
    \centering
    \includegraphics[width=0.7\linewidth]{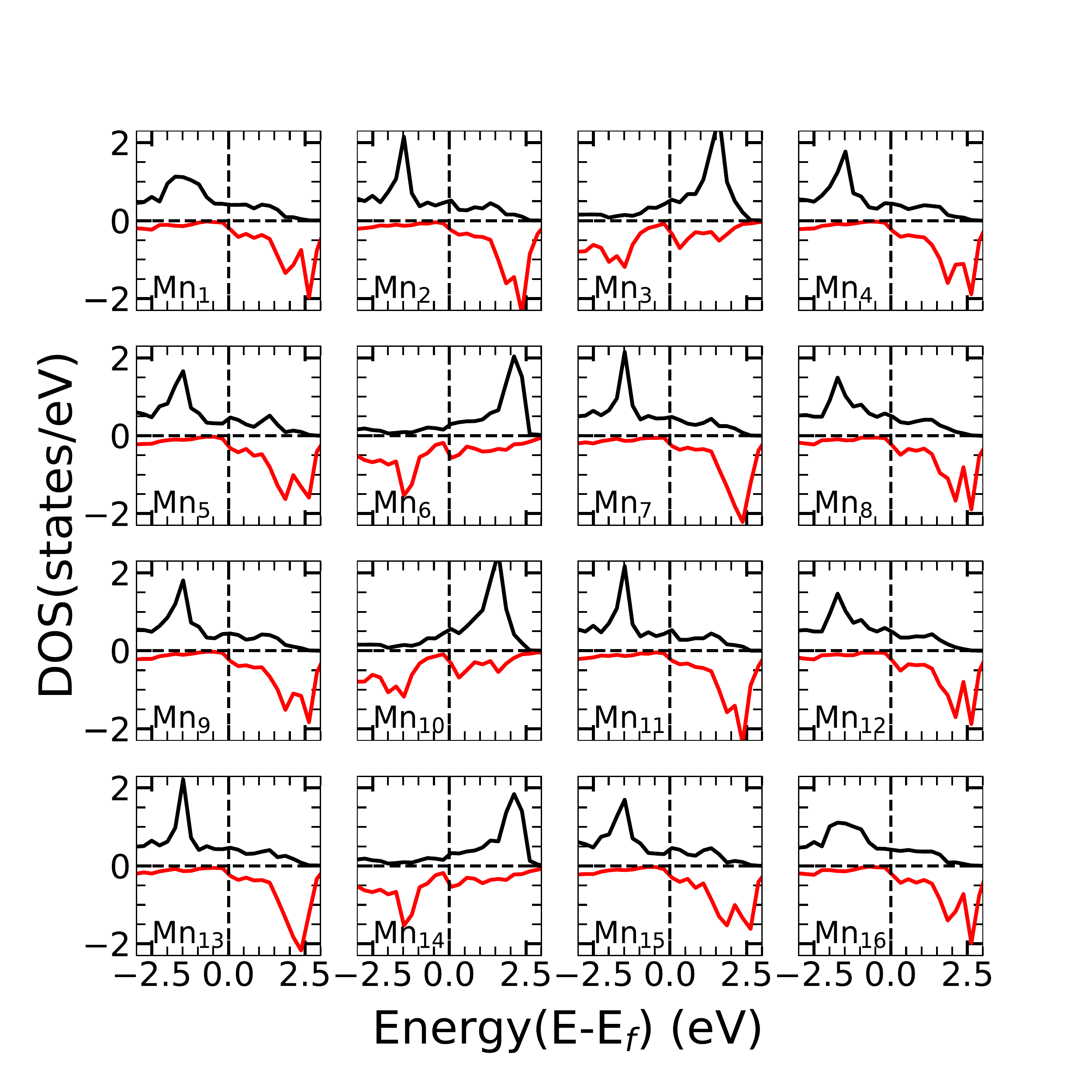}
    \caption{Projected Densities of States each Mn atoms in $\eta$=0.50 of V$_{1.0}$Mn$_{1.0}$CO$_2$}
    \label{fig:9}
\end{figure}

\begin{figure}[H]
    \centering
    \includegraphics[width=0.7\linewidth]{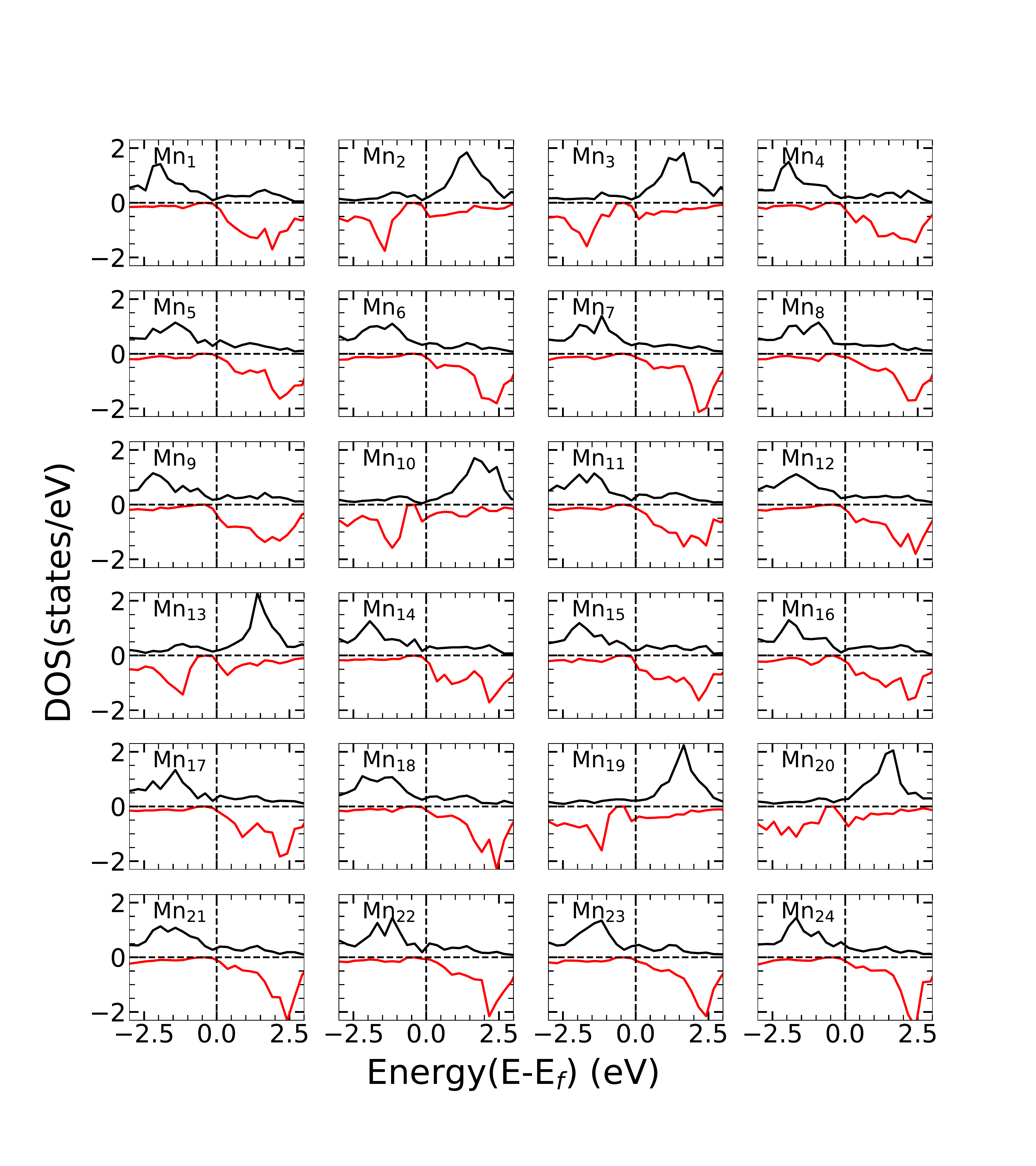}
    \caption{Projected Densities of States each Mn atoms in $\eta$=0.50 of V$_{0.5}$Mn$_{1.5}$CO$_2$}
    \label{fig:10}
\end{figure}

\begin{figure}[H]
    \centering
    \includegraphics[width=0.7\linewidth]{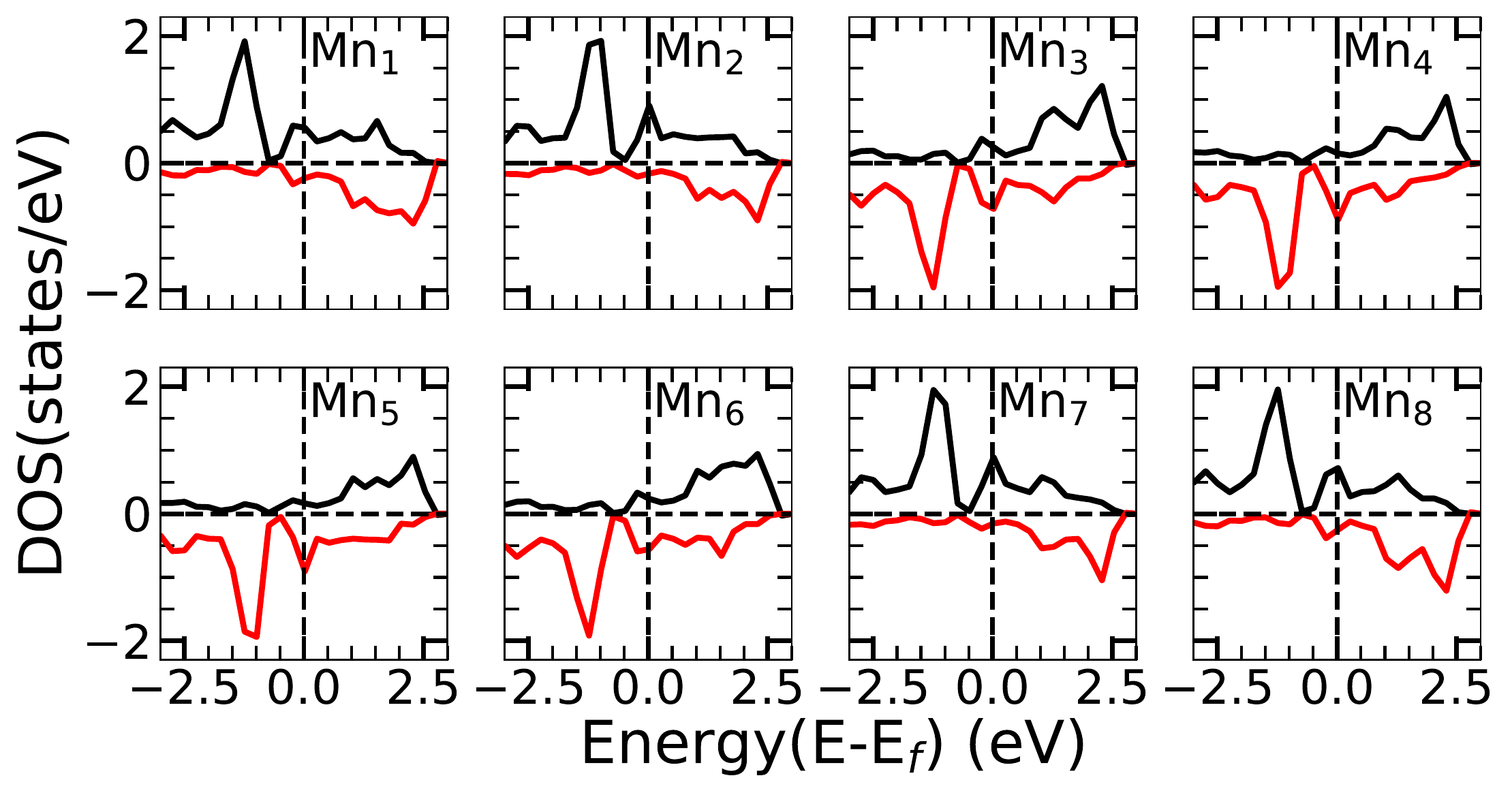}
    \caption{Projected Densities of States each Mn atoms in $\eta$=0.00 of V$_{1.5}$Mn$_{0.5}$CO$_2$}
    \label{fig:11}
\end{figure}

\begin{figure}[H]
    \centering
    \includegraphics[width=0.7\linewidth]{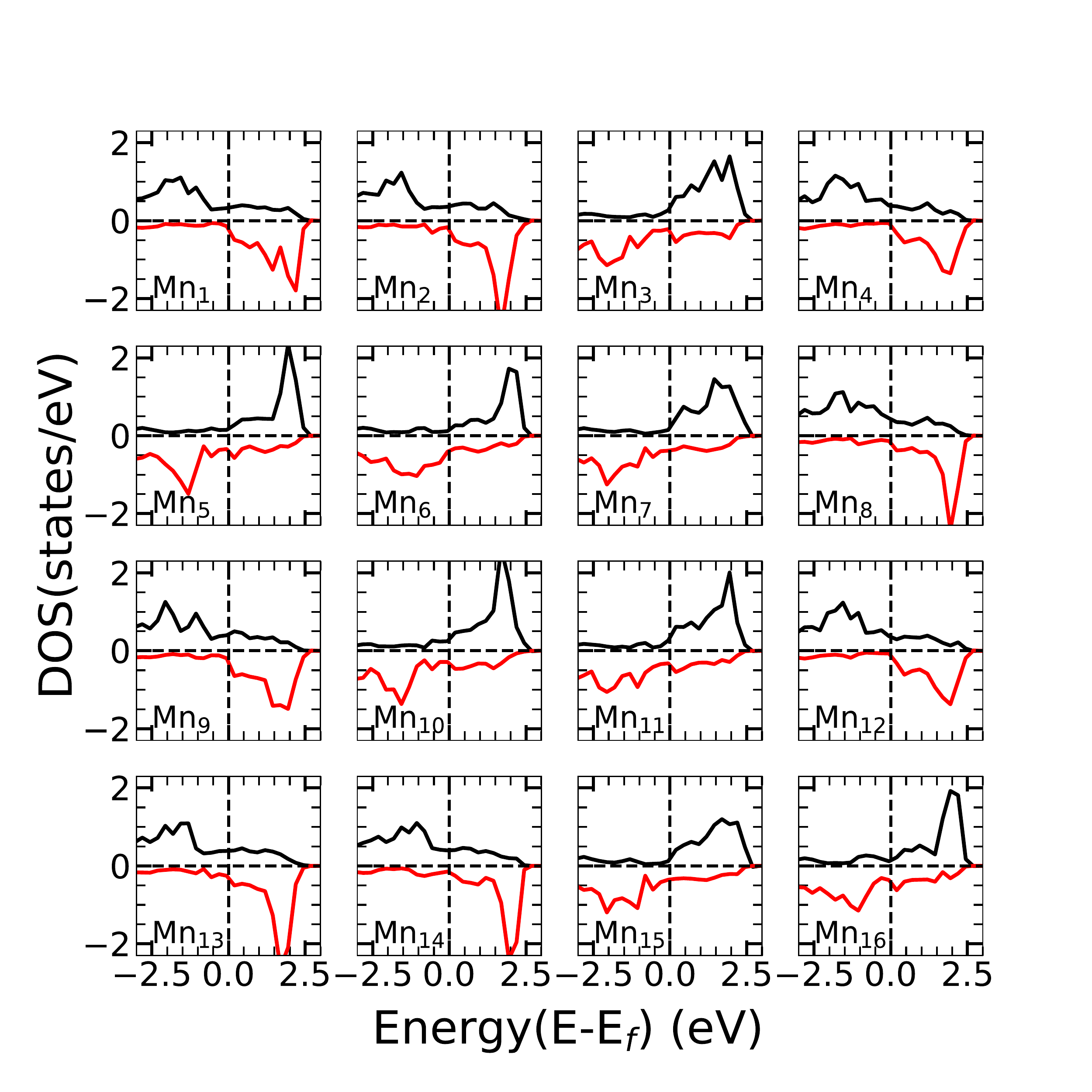}
    \caption{Projected Densities of States each Mn atoms in $\eta$=0.00 of V$_{1.0}$Mn$_{1.0}$CO$_2$}
    \label{fig:12}
\end{figure}

\begin{figure}[H]
    \centering
    \includegraphics[width=0.7\linewidth]{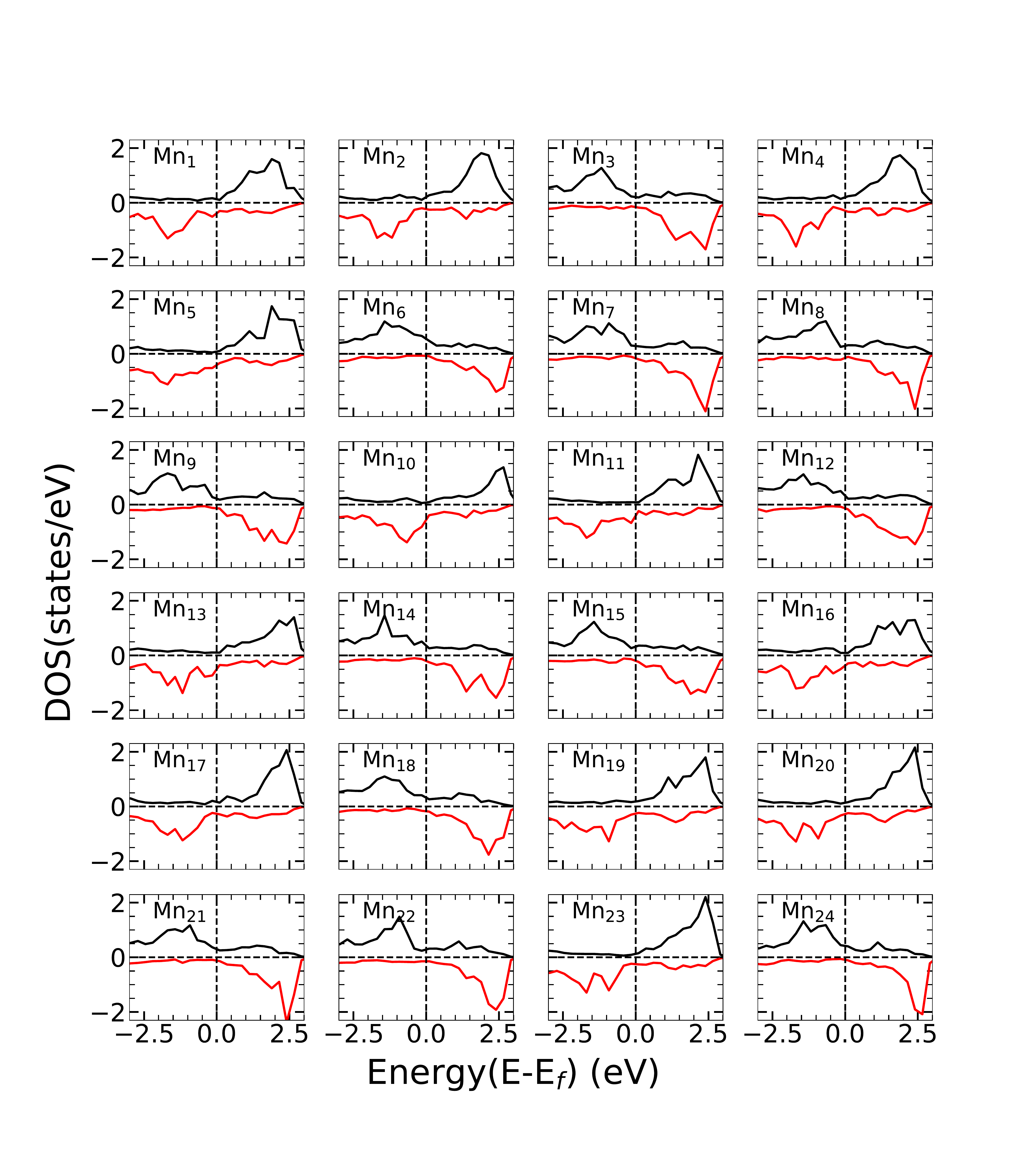}
    \caption{Projected Densities of States each Mn atoms in $\eta$=0.00 of V$_{0.5}$Mn$_{1.5}$CO$_2$}
    \label{fig:13}
\end{figure}

\begin{figure}[H]
    \centering
    \includegraphics[width=0.7\linewidth]{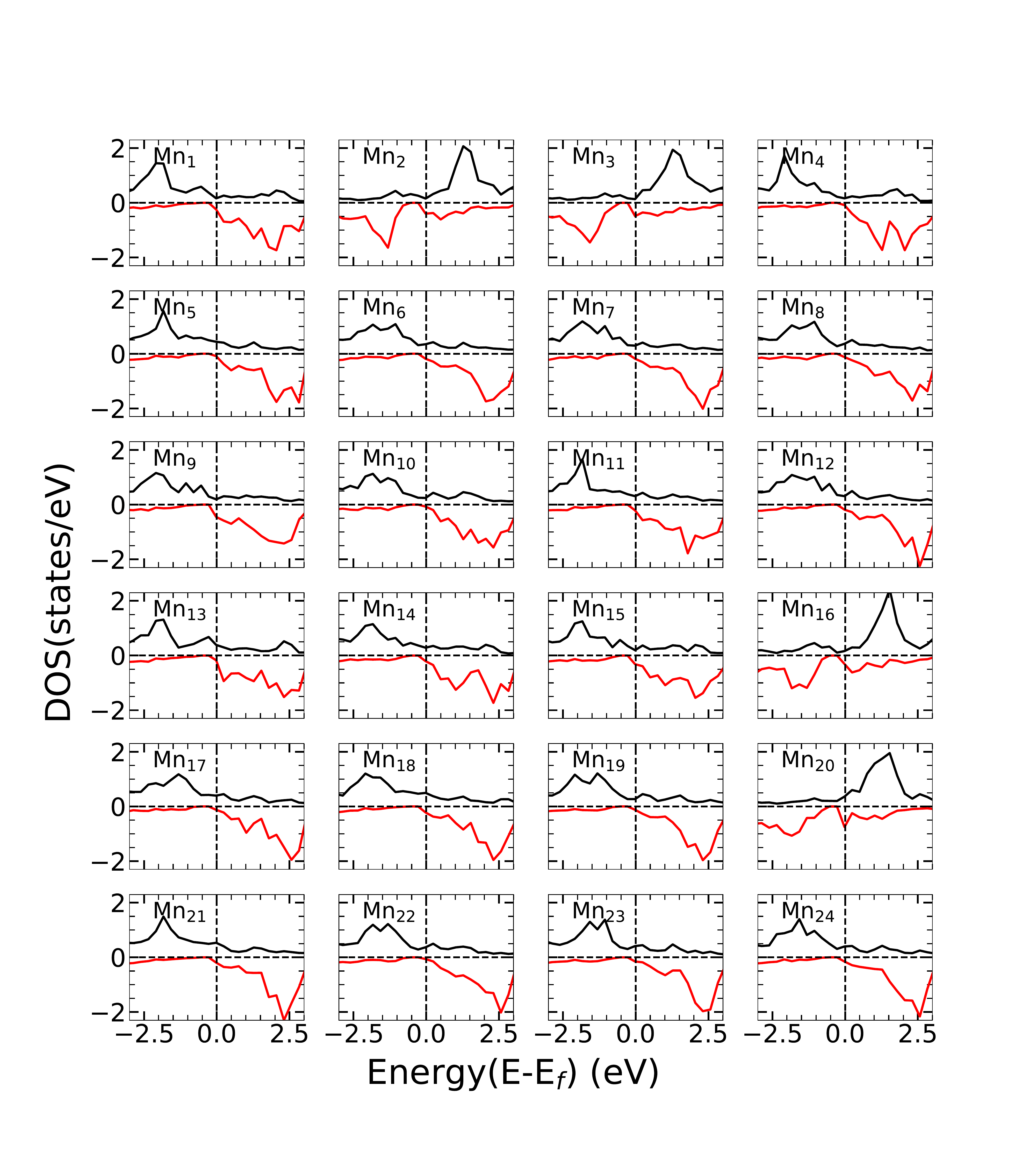}
    \caption{Projected Densities of States each Mn atoms in $\eta$=0.66 of V$_{0.5}$Mn$_{1.5}$CO$_2$}
    \label{fig:14}
\end{figure}

The figures \ref{fig:1} (a)-(b), (c)-(d) and (e)-(f) are the top view of top-bottom surfaces of the ground state of V$_{0.5}$Mn$_{1.5}$CO$_2$, V$_{1.0}$Mn$_{1.0}$CO$_2$ and V$_{1.5}$Mn$_{0.5}$CO$_2$, respectively. The side views and detailed descriptions of the same are represented in figure \ref{Fig:4} and section \ref{result-structure}, respectively, of the main text. Figure \ref{fig:2} shows projected densities of states of each Mn present in V$_{1.5}$Mn$_{0.5}$CO$_2$ in the ground state. The Mn present in this system shows anti-ferromagnetic ordering along the +c-axis. The top layer Mn, $i.e$, Mn$_{1}$-Mn$_{4}$ have a sharp peak in -2.0 $eV$ to 0 $eV$ range of occupied part suggests their spin moment along the +c-axis. The bottom layer Mn, $i.e$, Mn$_{5}$-Mn$_{8}$ have mirror image densities of states of Mn$_{1}$-Mn$_{4}$ suggesting anti-aligned spin moment with c-axis. This system is ferromagnetic along the a and b axis. In figure \ref{fig:3}, we represent the electronic structure of each vanadium present in V$_{1.5}$Mn$_{0.5}$CO$_2$. The presence of lesser states in the occupied part confirms 0.02 $\mu_B$ of a magnetic moment over them. $\eta$=0.00 state is a fully disordered local moment system, where 50$\%$ on Mn atoms have spin moment along the -c-axis in one surface. 
Figures \ref{fig:4} and \ref{fig:5} show Mn and V projected densities of states of each Mn and V of V$_{1.0}$Mn$_{1.0}$CO$_2$. Mn$_1$ to Mn$_8$ is the top layer, and Mn$_{9}$ to Mn$_{16}$ are the bottom layer of manganese in this system. As shown in Figure \ref{fig:4}, all Mn has most of the densities of states in the positive spin channel compared to the negative spin channel in the occupied part, confirming the Mn atoms' ferromagnetic ordering. All the V has fewer states in the occupied part (Figure \ref{fig:5}) and shows a magnetic moment of 0.05 $\mu_B$. Figure \ref{fig:6} shows the electronic structure of all Mn present in the system V$_{0.5}$Mn$_{1.5}$CO$_2$ in the ground state. As discussed in the main text, this system is special in ground-state magnetic ordering. This state exhibits ferri-magnetic ordering. All the Mn atoms of the top surface, Mn$_{1}$ to Mn$_{12}$, have higher states in the up spin channel in the occupied part exhibiting ferromagnetic ordering in the top surface. Mn$_{1}$ to Mn$_{4}$ have 2.7 $\mu_B$ and Mn$_{5}$ to Mn$_{12}$ have 2.6 $\mu_B$ local moment of them, respectively. Four Mn atoms (Mn$_{13}$ to Mn$_{16}$) on the bottom surface have an anti-aligned spin direction to the other eight Mn atoms (Mn$_{17}$ to Mn$_{24}$) of that surface. This makes the bottom surface ferrimagnetic in order. The anti-aligned Mn has a local spin moment of 1.7 $\mu_B$, and the other aligned Mn has a 2.6 $\mu_B$ moment.
Figure \ref{fig:7} shows the densities of states of eight vanadium atoms present in the system V$_{0.5}$Mn$_{1.5}$CO$_2$. The vanadium in the top surface V$_{1}$ to V$_{4}$ have local moment 0.38 $\mu_B$ and bottom surface V$_{5}$ to V$_{8}$ have 0.14 $\mu_B$.
In figures \ref{fig:8}, \ref{fig:9} and \ref{fig:10}, we have represented the Mn projected densities of states of V$_{2-x}$Mn$_{x}$CO$_2$ for x=0.5, 1.0 and 1.5 respectively in $\eta$=0.50 states. In $\eta$=0.50 state, 25$\%$ of Mn atoms have spin direction aligned anti-parallel to the c-axis in one surface. Mn$_{4}$ in the top surface and Mn$_{5}$ in the bottom surface have spin direction in the -c-direction for x=0.5 system (Figure \ref{fig:8}). For V$_{1.0}$Mn$_{1.0}$CO$_2$, Mn$_3$, Mn$_6$ of top surface and Mn$_{10}$, Mn$_{14}$ of bottom surface have local spin direction anti-parallel to the +c-axis (Figure \ref{fig:9}). For the Mn-rich system(x=1.5), Mn$_{2}$, Mn$_{3}$, Mn$_{10}$ of top surface and Mn$_{13}$, Mn$_{19}$ and Mn$_{20}$ of bottom surface have spin in the -c-direction(Figure \ref{fig:10}).The figures \ref{fig:11}, \ref{fig:12}, \ref{fig:13}, we have shown the Mn projected densities of states for all chemical disorder compound V$_{2-x}$Mn$_{x}$CO$_2$, x = 0.5, 1.0, 1.5. Figure \ref{fig:14} shows the densities of states of each Mn atom of V$_{0.5}$Mn$_{1.5}$CO$_2$ in the $\eta$=0.66 state. The reason for highlighting this state is an equal number of aligned and anti-aligned spin Mn atoms as the ground state in this $\eta$=0.66 state. This state also has  20 Mn atoms with the spin direction along +c-axis and 4 Mn atoms along the -c-axis. In this system, anti-aligned Mn atoms are present on both surfaces, unlike the ground state system.

\section{Anomalous charge transfer of V$_{0.5}$Mn$_{1.5}$CO$_2$}
\begin{table}[ht!]
\setlength{\tabcolsep}{8pt} 
\renewcommand{\arraystretch}{1.0} 
    \caption{Charge transfer from each surface of V$_{2-x}$Mn$_{x}$CO$_2$ for x=0.5, 1.0 and 1.5 in the ground and all other $\eta$ state}
    \begin{tabular}{|c@{\hspace{0.2cm}} |c@{\hspace{0.2cm}}| c|c@{\hspace{0.2cm}}|} 
    \hline
    System & States & \multicolumn{2}{c|}{surface contribution} \\
    \hline
               &    &    Top & Bottom  \\  
    \hline
    V$_{1.5}$Mn$_{0.5}$CO$_{2}$ & GS(AFM-c) & 7.095$e$ & 7.095$e$  \\
                                & $\eta$ =0.50 & 7.025$e$ & 7.025$e$\\
                                & $\eta$ =0.00 & 7.045$e$ & 7.045$e$ \\
    \hline
    V$_{1.0}$Mn$_{1.0}$CO$_{2}$ & GS(FM) & 7.125$e$ & 7.125$e$  \\
                                & $\eta$=0.75 & 7.13$e$ & 7.130$e$\\
                                & $\eta$=0.50 &  7.135$e$ &  7.135$e$ \\
                                & $\eta$=0.25 &  7.14$e$  &  7.140$e$ \\
                                & $\eta$=0.00 &  7.145$e$ & 7.145$e$  \\
    \hline                            
    V$_{0.5}$Mn$_{1.5}$CO$_{2}$ & GS(FeM) & 9.648$e$ & 6.352$e$ \\
                                & $\eta$ =0.66 & 7.065$e$ & 7.065$e$  \\
                                & $\eta$ =0.50 & 7.07$e$ & 7.070$e$  \\
                                & $\eta$ =0.33 & 7.06$e$ & 7.060$e$  \\
                                & $\eta$ =0.16 & 7.07$e$ & 7.070$e$  \\
                                & $\eta$ =0.00 & 7.11$e$ & 7.110$e$\\
    \hline
    \end{tabular}
    \label{tab:1}
\end{table}    

\begin{figure}[H]
    \centering
    \includegraphics[width=0.7\linewidth]{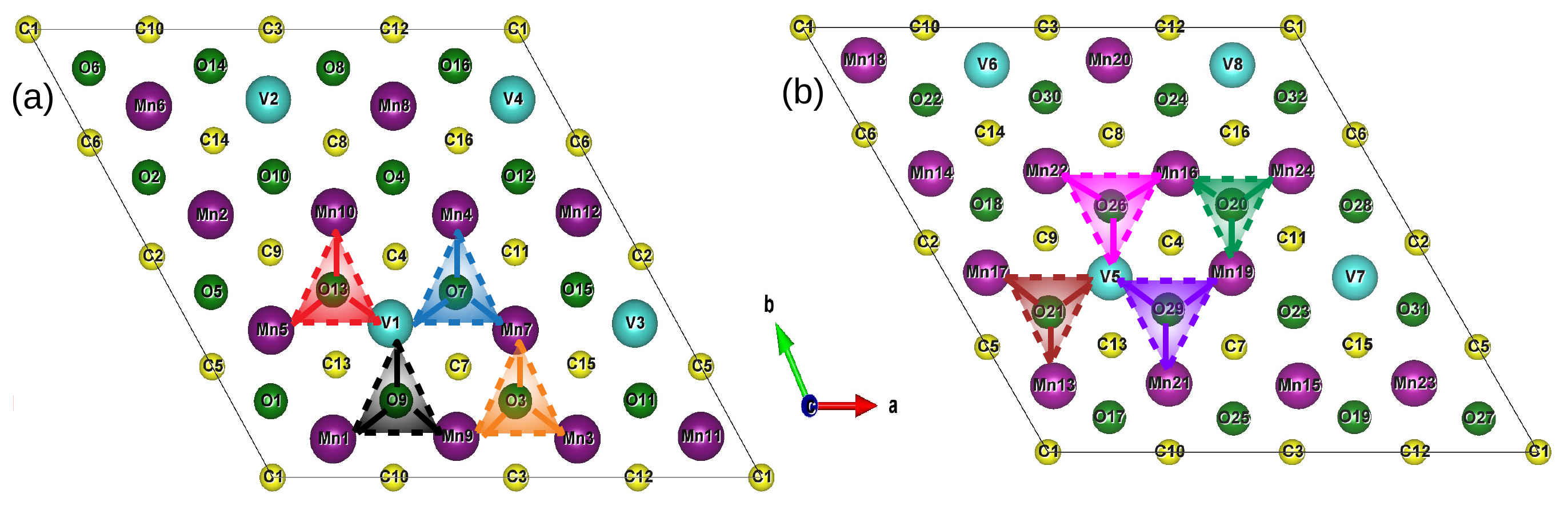}
    \caption{(a) top layer and (b) bottom layer of  V$_{0.5}$Mn$_{1.5}$CO$_{2}$ }
    \label{fig:15}
\end{figure}

In Table \ref{tab:1}, we have shown charge transfer corresponding to each surface of V$_{2-x}$Mn$_{x}$CO$_2$, x=0.5, 1.0, 1.5 in ground and all other $\eta$ states.
To understand the unequal charge transfer between the electrode and hydrogen from the two surfaces of V$_{0.5}$Mn$_{1.5}$CO$_2$ in the ground state, we have shown individual charge on each element before(no-H) and after(with-H) hydrogenation in tables \ref{tab:2}(top) and \ref{tab:3}(bottom). As the hydrogen interacts with the electrode through the oxygen, we have performed a local environment study with respect to oxygen. While studying the local environment, it is observed that 16 oxygen can be divided into four groups. The environment of one oxygen from each group is shown in Figure \ref{fig:15}(a)(top) and the charge on that oxygen and the nearest neighbor transition elements are shown in Table \ref{tab:2}. Oxygen atoms belonging to a group have the same charge and local environment before and after H adsorption. For example, O$_1$, O$_2$, O$_3$, and O$_4$ have the same environment but we have highlighted the environment of O$_3$ only with orange color in Figure \ref{fig:15}(a) and Table \ref{tab:2} shows the corresponding charges. Accordingly, O$_{5-8}$, O$_{9-12}$, O$_{13-16}$ belong to the groups 2, 3, and 4, respectively, and O$_7$, O$_9$, and O$_{13}$ is highlighted from each group in Figure \ref{fig:15}(a). Similarly Table \ref{tab:3} and Figure \ref{fig:15}(b) represent the local environment and charge information for all oxygen atoms in the bottom layer of V$_{0.5}$Mn$_{1.5}$CO$_2$. 

\begin{table}[H]
    \caption{Charge on each element of V$_{0.5}$Mn$_{1.5}$CO$_2$(top layer) before(no-H) and after(with-H) hydrogenation. The distance of oxygen from the nearest neighbor transition metal atom is also given.}
    \centering
    \begin{tabular}{|c|c|c|c|c|}
    \hline
    Surface & Group & No-H & With-H & Distance\\
    \hline
            &       & charge ($e$) & charge ($e$) & \\
    \hline
    \multirow{20}{*}{Top}   &  \multirow{5}{*}{1(orange)}   &  O$_{3}$(6.91) & O$_{3}$(7.31) & - \\
    \cline{3-5}
          &               & Mn$_{3}$(11.58)& Mn$_{3}$(11.65)  & O$_{3}$-Mn$_{3}$(1.95) \\
    \cline{3-5}
          &               & Mn$_{7}$(11.60) & Mn$_{7}$(11.68)&  O$_{3}$-Mn$_{7}$(1.92) \\
    \cline{3-5}
          &               & Mn$_{9}$(11.60) & Mn$_{9}$(11.65) &  O$_{3}$-Mn$_{9}$(1.92)  \\ 
    \cline{3-5}
          &               & - & H$_{3}$(0.45) &  -  \\  
    \cline{2-5}   
          &  \multirow{5}{*}{2(blue)}   &  O$_{7}$(6.94) & O$_{7}$(7.34) & -  \\
    \cline{3-5}
          &               & Mn$_{4}$(11.58) & Mn$_{4}$(11.65) & O$_{7}$-Mn$_{4}$(2.00) \\
    \cline{3-5}
          &               & Mn$_{7}$(11.60) & Mn$_{7}$(11.68) & O$_{7}$-Mn$_{7}$(1.99)  \\
    \cline{3-5}
          &               & V$_{1}$(11.07) & V$_{1}$(11.42) &  O$_{7}$-V$_{1}$(1.81)   \\
    \cline{3-5}
          &               & - & H$_{7}$(0.42) &  -  \\ 
    \cline{2-5}    
          &  \multirow{5}{*}{3(black)}   &  O$_{9}$(6.94) &  O$_{9}$(7.32)& -\\
    \cline{3-5}
          &               & Mn$_{1}$(11.58) & Mn$_{1}$(11.65) & O$_{9}$-Mn$_{1}$(1.98) \\
    \cline{3-5}
          &               & Mn$_{9}$(11.60) & Mn$_{9}$(11.65) & O$_{9}$-Mn$_{9}$(1.97)  \\
    \cline{3-5}
          &               & V$_{1}$(11.07) & V$_{1}$(11.42) &  O$_{9}$-V$_{1}$(1.83)   \\
    \cline{3-5}
          &               & - & H$_{9}$(0.45) &  -  \\ 
   \cline{2-5}    
         &  \multirow{5}{*}{4(red)}  &  O$_{13}$(6.88) & O$_{13}$(7.40) & - \\
    \cline{3-5}
          &               & Mn$_{5}$(11.60) & Mn$_{5}$(11.68) & O$_{13}$-Mn$_{5}$(2.10) \\
    \cline{3-5}
          &               & Mn$_{10}$(11.60) & Mn$_{10}$(11.65) & O$_{13}$-Mn$_{10}$(2.04)  \\
    \cline{3-5}
          &               & V$_{1}$(11.07) &   V$_{1}$(11.07) & O$_{13}$-V$_{1}$(1.72)   \\
    \cline{3-5}
          &               & - & H$_{13}$(0.27) &  -  \\ 
    \hline \hline    
    \end{tabular}
    \label{tab:2}
\end{table}

\begin{table}[H]
    \caption{Charge on each element of V$_{0.5}$Mn$_{1.5}$CO$_2$(bottom layer) before(no-H) and after(with-H) hydrogenetion. The distance of oxygen from the nearest neighbor transition metal atom is also given.}
    \centering
    \begin{tabular}{|c|c|c|c|c|}
    \hline
    Surface & Group & No-H & With-H & Distance\\
    \hline
            &       & charge & charge & \\
    \hline
    \multirow{16}{*}{Bottom}   &  5(green)   &  O$_{20}$(6.91) & O$_{20}$(7.24) & - \\
    \cline{3-5}
          &               & Mn$_{16}$(11.68)& Mn$_{16}$(11.68)  & O$_{20}$-Mn$_{16}$(1.93) \\
    \cline{3-5}
          &               & Mn$_{19}$(11.60) & Mn$_{19}$(11.64)&  O$_{20}$-Mn$_{19}$(1.92) \\
    \cline{3-5}
          &               & Mn$_{24}$(11.58) & Mn$_{24}$(11.64) &  O$_{20}$-Mn$_{24}$(1.92)  \\  
    \cline{3-5}
          &               & - & H$_{20}$(0.51) &  -  \\  
    \cline{2-5}   
          &  6(brown)   &  O$_{21}$(6.93) & O$_{21}$(7.10) & -  \\
    \cline{3-5}
          &               & Mn$_{13}$(11.68) & Mn$_{13}$(11.68) & O$_{21}$-Mn$_{13}$(2.06) \\
    \cline{3-5}
          &               & Mn$_{17}$(11.60) & Mn$_{17}$(11.64) & O$_{21}$-Mn$_{17}$(1.99)  \\
    \cline{3-5}
          &               & V$_{5}$(11.07) & V$_{1}$(11.31) &  O$_{21}$-V$_{5}$(1.78)   \\
    \cline{3-5}
          &               & - & H$_{21}$(0.69) &  -  \\  
          
    \cline{2-5}    
          &  7(magenta)   &  O$_{26}$(6.91) &  O$_{26}$(7.10)& -\\
    \cline{3-5}
          &               & Mn$_{16}$(11.68) & Mn$_{16}$(11.68) & O$_{26}$-Mn$_{16}$(2.04) \\
    \cline{3-5}
          &               & Mn$_{22}$(11.58) & Mn$_{22}$(11.64) & O$_{26}$-Mn$_{22}$(1.97)  \\
    \cline{3-5}
          &               & V$_{5}$(11.07) & V$_{5}$(11.31) &  O$_{26}$-V$_{5}$(1.80)   \\
    \cline{3-5}
          &               & - & H$_{26}$(0.69) &  -  \\  
   \cline{2-5}    
         &  8(violet)   &  O$_{29}$(6.90) & O$_{29}$(7.20) & - \\
    \cline{3-5}
          &               & Mn$_{19}$(11.60) & Mn$_{19}$(11.64) & O$_{29}$-Mn$_{19}$(2.05) \\
    \cline{3-5}
          &               & Mn$_{21}$(11.58) & Mn$_{21}$(11.64) & O$_{29}$-Mn$_{21}$(2.04)  \\
    \cline{3-5}
          &               & V$_{5}$(11.07) &   V$_{5}$(11.31) & O$_{29}$-V$_{5}$(1.75)   \\
    \cline{3-5}
          &               & - & H$_{29}$(0.51) &  -  \\  
    \hline \hline    
    
    \end{tabular}
    \label{tab:3}
\end{table}

\end{appendix}

\end{document}